\documentclass[notitlepage]{revtex4-1}

\usepackage[utf8]{inputenc}
\usepackage[colorlinks=true,citecolor=blue,linkcolor=blue]{hyperref}
\usepackage[normalem]{ulem}
\usepackage{url}
\usepackage{graphicx,wrapfig,float,slashed,cancel}
\usepackage{amsmath,amssymb,epsfig,graphicx,xcolor,stmaryrd}
\usepackage{bm}
\usepackage{enumitem}
\usepackage{multirow}

\definecolor{darkblue}{RGB}{1, 90, 173}

\begin{document}

\title{Investigation of hidden-charm double strange pentaquark candidate $P_{css}$ via its mass and strong decays}

\author{K.~Azizi}
\email{ kazem.azizi@ut.ac.ir}
\thanks{Corresponding author}
\affiliation{Department of Physics, University of Tehran, North Karegar Avenue, Tehran
14395-547, Iran}
\affiliation{Department of Physics, Do\v{g}u\c{s} University, Dudullu-\"{U}mraniye, 34775
Istanbul, Turkey}
\author{Y.~Sarac}
\email{yasemin.sarac@atilim.edu.tr}
\affiliation{Electrical and Electronics Engineering Department,
Atilim University, 06836 Ankara, Turkey}
\author{H.~Sundu}
\email{ hayriye.sundu@kocaeli.edu.tr}
\affiliation{Department of Physics, Kocaeli University, 41380 Izmit, Turkey}

\date{\today}

\preprint{}

\begin{abstract}

This work presents an analysis of a hidden charmed pentaquark candidate state with double strange quark in its quark content. The investigation is performed in two parts, which provide the mass prediction of the considered state and its partial decay widths for the $P_{css}\rightarrow \Xi^0 J/\psi$ and $P_{css}\rightarrow \Xi_c^+ D_s^-$ channels. For the analyses, two-point and three-point QCD sum rule methods are applied to get the mass and the widths, respectively. The mass for this candidate state is obtained as $m_{P_{css}}=4600 \pm 155$~MeV with corresponding current coupling constant $\lambda_{P_{css}}=(0.81 \pm 0.21)\times 10^{-3}$~GeV$^6$. These results are used in the analysis of the partial widths of this state for the decays $P_{css}\rightarrow \Xi^0 J/\psi$ and $P_{css}\rightarrow \Xi_c^+ D_s^-$. From these decays, the total width is obtained as $\Gamma = 12.29\pm 2.94 $~MeV. These results may shed light on the future experimental searches in which such types of states are probed and may provide information to discriminate between such possible observations.
   
\end{abstract}


\maketitle

\renewcommand{\thefootnote}{\#\arabic{footnote}}
\setcounter{footnote}{0}
\section{\label{sec:level1}Introduction}\label{intro} 

In the last few decades, the  significant progress in experiments has brought new particles to light. With their quark substructure, some of these newly observed particles differ from conventional hadrons, which are composed of quark-antiquark (mesons) or three quarks (baryons). Quarks are combined in color singlet states to form these hadrons, and nature does not rule out the non-conventional hadrons, which consist of other multiquark combinations in color singlet forms such as tetraquarks and pentaquarks. This fact has always attracted interests toward these non-conventional states. They were firstly proposed by Gell-Mann~\cite{Gell-Mann}, and their existences and properties have been investigated thoroughly since then from different points of view using different approaches in both experiments and theory. Their first observation was announced in 2003 by Belle Collaboration for a tetraquark state, $X(3872)$~\cite{Choi:2003ue}. This observation was later confirmed by other  collaborations such as the BABAR Collaboration~\cite{BaBar:2004oro},CDF II Collaboration~\cite{CDF:2003cab,CDF:2009nxk}, D0 Collaboration~\cite{D0:2004zmu}, LHCb Collaboration~\cite{LHCb:2011zzp}, and CMS Collaboration~\cite{CMS:2013fpt}. Observation of other tetraquark states followed this observation.

On the other hand,  hidden charm pentaquark states, which are other members of these exotic states, were firstly observed in 2015 by the LHCb Collaboration~\cite{Aaij:2015tga} in the $J/\psi p$ invariant mass spectrum of the $\Lambda_b^0 \rightarrow J/\psi p K^-$ process. The resonance properties for these pentaquark states were reported as $m_{P_c(4380)^+}=4380 \pm8 \pm 29~\mathrm{MeV}$, $\Gamma_{P_c(4380)^+}=205 \pm 18 \pm 86~\mathrm{MeV}$ and $m_{P_c(4450)^+}=4449.8 \pm 1.7 \pm 2.5~\mathrm{MeV}$, $\Gamma_{P_c(4450)^+}= 39 \pm 5 \pm 19~\mathrm{MeV}$~\cite{Aaij:2015tga}. In 2016, full amplitude analysis for $\Lambda_b^0 \rightarrow J/\psi p \pi^-$ decays~\cite{Aaij:2016ymb} supported this observation. And later, in 2019, the LHCb collaboration  updated the analysis with a combined data set and confirmed the previously observed states, and also observed additional structures which became more significant with additional data set compared to the previous analysis~\cite{Aaij:2019vzc}. The resonance parameters for the additionally observed state, $P_c(4312)^+$, were presented as $m_{P_c(4312)^+}=4311.9 \pm 0.7^{ +6.8}_{-0.6}~\mathrm{MeV}$ and $\Gamma_{P_c(4312)^+}=9.8 \pm 2.7 ^{ +3.7}_{-4.5}~\mathrm{MeV}$~\cite{Aaij:2019vzc}. In this analysis, for the previously observed peak of $P_c(4450)^-$, there occurred a split into two peaks with the following mass and width: $m_{P_c(4440)^+}=4440.3 \pm 1.3 ^{+ 4.1}_{-4.7}~\mathrm{MeV}$, $\Gamma_{P_c(4440)^+}= 20.6 \pm 4.9^{+8.7}_{-10.1}~\mathrm{MeV}$ and $m_{P_c(4457)^+}=4457.3 \pm 0.6 ^{+ 4.1}_{-1.7}~\mathrm{MeV}$, $\Gamma_{P_c(4457)^+}= 6.4 \pm 2.0^{+5.7}_{-1.9}~\mathrm{MeV}$~\cite{Aaij:2019vzc}. The next observation consistent with possible pentaquark state came with a strange quark in its quark content~\cite{LHCb:2020jpq}, which was observed in the $J/\psi \Lambda$ invariant mass distribution analysis of the $\Xi_b^- \rightarrow J/\psi \Lambda K^-$ decays and denoted as $P_{cs}(4459)$, with mass $m_{P_{cs}(4459)}=4458.8\pm 2.9^{+4.7}_{-1.1}$~MeV and width $17.3\pm 6.5^{+8.0}_{-5.7}$~MeV. Very recently the LHCb Collaboration announced an evidence for  a pentaquark candidate state in the $B_s^0\rightarrow J/\psi p \bar{p}$ decays with mass and width $m_{P_{c}(4337)}=4337^{+7}_{-4}{}^{+2}_{-2}$~ MeV and $\Gamma_{P_{c}(4337)}=29^{+26}_{-12}{}^{+14}_{-14}$~ MeV, respectively~\cite{LHCb:2021chn}.

Though the possible pentaquark states have been investigated before all these observations, with these experimental findings, they have recaptured the theoretical attentions over themselves again, and they were extensively investigated for their different properties using various theoretical approaches. Their spectroscopy and interactions were examined to explain their substructures and quantum numbers. Among the considered possible substructures is the molecular interpretation suggested based on their closeness to the meson baryon thresholds. The molecular structure was taken into account in the one-pion-exchange model~\cite{Chen:2016heh}, in the QCD sum rules method~\cite{Chen:2016otp,Azizi:2016dhy,Azizi:2018bdv,Azizi:2020ogm}, in the quasipotential Bethe-Salpeter equation approach~\cite{He:2019ify,Zhu:2021lhd}, in the contact-range effective field theory approach~\cite{Liu:2019tjn}, in the effective Lagrangian approach~\cite{Xiao:2019mvs,Lu:2016nnt}, in the constituent chiral quark model~\cite{Hu:2021nvs} and in the chiral perturbation theory~\cite{Meng:2019dba}. The meson baryon molecular interpretation was also adopted in Refs.~\cite{Chen:2015loa,Chen:2015moa,He:2015cea,Meissner:2015mza,Roca:2015dva,Chen:2020opr,Yan:2021nio,Wu:2021caw,Chen:2020uif,Phumphan:2021tta,Du:2021fmf,Lu:2021irg,Gao:2021hmv,Yalikun:2021dpk}. Ref.~\cite{Chen:2021obo} used the dynamically constrained phase space coalescence model and the parton and hadron cascade model and studied three kinds of possible structures  for the $P_c$ states, which are the pentaquark state, the nucleus-like state, and the molecular state.  These states were also considered to be compact pentaquark states. In Refs.~\cite{Wang:2015epa,Maiani:2015vwa,Giannuzzi:2019esi,Li:2015gta,Lebed:2015tna,Anisovich:2015cia,Wang:2015ava,Wang:2015ixb,Ghosh:2015ksa,Wang:2015wsa,Zhang:2017mmw,Wang:2019got,Wang:2020rdh,Ali:2020vee,Wang:2016dzu,Azizi:2021utt,Anisovich:2015zqa,Wang:2020eep,Gao:2021hmv}, the substructures of the pentaquark states were treated to be diquark-diquark-antiquark forms. Another possible substructure, diquark-triquark, was also investigated in Refs.~\cite{Wang:2016dzu,Zhu:2015bba,Gao:2021hmv}. The possibility of their being kinematical effect was regarded in Refs.~\cite{Guo1,Guo2,Mikhasenko:2015vca,Liu1,Bayar:2016ftu,Nakamura:2021qvy}. In Ref.~\cite{Nakamura:2021dix}, the newly observed  $P_{c}(4337)$ was considered by developing a model to analyze the data on $B_s^0\rightarrow J/\psi p \bar{p}$.  Despite all these works, both the substructure and quantum numbers of these observed states are still not certain and they need further investigations both theoretically and experimentally for their clarifications.  While many works were devoted to explain the properties of the observed pentaquark states, others were carrying the aim of offering new possible pentaquark states with different quark contents to be observed in future experimental investigations. Among these works are the ones given in Refs.~\cite{Liu:2020cmw,Chen:2015sxa,Feijoo:2015kts,Irie:2017qai,Chen:2016ryt,Zhang:2020cdi,Paryev:2020jkp,Gutsche:2019mkg,Azizi:2017bgs,Azizi:2018dva,Cao:2019gqo,Zhang:2020vpz,Xie:2020ckr,Meng:2019fan,Liu:2021tpq,Wang:2020bjt,Wang:2019nvm,Liu:2021ixf,Giron:2021fnl,Santopinto:2016pkp,Xiao:2019gjd}. 

There are many works devoted to the investigation of these states. However, their properties are still not certain, and there are ambiguities in their substructures and the quantum numbers as well. Different works suggested different interpretations for the properties of the  observed pentaquark states. To clarify all these ambiguities, we need more investigations on the properties of both the observed states and the new possible ones to be observed. In these investigations, either complementary reactions or some other decay modes can be considered for the presently observed pentaquark states, or the new particles that are possible to be observed can be probed, to provide inputs to the experiments, by their spectroscopic properties or possible decay modes. The analyses for such processes can be found in Refs.~\cite{Feijoo:2015kts,Liu:2021ixf,Chen:2015sxa,Wang:2015jsa,Kubarovsky:2015aaa,Karliner:2015voa,Cheng:2015cca,Liu:2020ajv,Yang:2021pio,Stancu:2021rro,Dong:2020nwk,Voloshin:2019aut,Ling:2021lmq,Xing:2021yid,Cheng:2021gca,Wang:2019hyc,Xu:2019zme,Liu:2021ojf}.

It is natural to expect  observations of  new pentaquark states, with different quark contents than the observed ones, in the near future. So far, with this expectation, some of such states were investigated. For instance, the pentaquark states with bottom quark were probed in Refs.~\cite{Chen:2016heh,Liu:2017xzo,Azizi:2017bgs,Huang:2021ave,Zhu:2020vto,Ferretti:2018ojb,Shimizu:2016rrd}. The pentaquark candidates with single or triple charm quark were investigated in Ref.~\cite{Azizi:2018dva} with a molecular interpretation. With the expectation to observe the fully-heavy pentaquark states, these types of pentaquark structures were considered in Refs.~\cite{Zhang:2020vpz,Wang:2021xao,An:2020jix,Yan:2021glh}. Another possible candidate for such states is the ones including strange quarks. The pentaquark with double strange quark was examined in Refs.~\cite{Wang:2020bjt} with molecular interpretation using the one-boson-exchange model. In Ref.~\cite{Meng:2019fan}, the pentaquark with triple strange quarks was considered. In Refs.~\cite{Ferretti:2020ewe,Ferretti:2021zis} the hadro-quarkonium model was used to search for the double strange pentaquarks. The pentaquarks with triple strange quark were analyzed in Ref.~\cite{Wang:2021hql} in the meson-baryon molecular picture via  the one-boson exchange model. In this respect, theoretical works involving the predictions for possible new pentaquark states may shed light on experimental investigations. With this motivation, in this work, we investigate a possible hidden charm pentaquark state involving double strange quarks. To this end, we apply the QCD sum rules method~\cite{Shifman:1978bx,Shifman:1978by,Ioffe81} which is among the powerful methods, and provided consistent predictions so far with the observed hadrons. In the application of the method, the main input is the proper interpolating field for the considered state. For the present study, we work with an interpolating field in the diquark-diquark-antiquark form and with $J^P=\frac{1}{2}^{-}$ spin-parity quantum numbers. Using the two-point QCD sum rule analysis, we first calculate the mass and current coupling constant of the candidate pentaquark state. Then we use these findings in the calculation of the widths of its strong decays to $\Xi^0 J/\psi$ and $\Xi_c^+ D_s^-$. 

The outline of the article is as follows: In Sec.~\ref{II} we present the two-point QCD sum rule analysis to calculate the mass and current coupling constant of the considered state (in what follows it is labeled as $P_{css})$ and present the analyses of the obtained results. Sec.~\ref{III} and Sec.~\ref{IV} are devoted to the calculation of the strong decay widths of the $P_{css}$ state to the states $\Xi^0 J/\psi$ and $\Xi_c^+ D_s^-$ via the application of three-point QCD sum rules, respectively. The last section presents the summary and conclusion.

\section{The mass of $P_{css}$ state}\label{II}

In this section, the mass of the candidate double strange pentaquark state is obtained. The  starting  point is to consider  the following two-point correlation function:
\begin{equation}
\Pi(q)=i\int d^{4}xe^{iq\cdot
x}\langle 0|\mathcal{T} \{\eta_{P_{css}}(x)\bar{\eta}_{P_{css}}(0)\}|0\rangle,
\label{eq:CorrFmassPc}
\end{equation}
where  $\mathcal{T}$ is the time ordering operator and $\eta^{P_{css}}$ is the interpolating current for the candidate pentaquark state $P_{css}$. In this work, we use an interpolating current in the diquark-diquark-antiquark form with spin-$\frac{1}{2}$ and negative parity. The interpolating current has the following explicit form:
\begin{eqnarray}
\eta_{P_{css}}&=&\epsilon^{ila}\epsilon^{ijk}\epsilon^{lmn}u^{T}_{j}C\gamma_5 s_k s^{T}_m C\gamma_5 c_{n} C \bar{c}^{T}_a.\label{CurrentPcss}
\end{eqnarray}
The interpolating current is constructed considering the valance quark content and the assumed quantum numbers of the state. $u$, $s$, and $c$ in the interpolating current correspond to the respective quark fields, and $C$ is the charge conjugation matrix.

The two-point correlation function is calculated following two paths. The first path gives its result in terms of hadronic degrees of freedom, such as the mass and current coupling constant of the considered state. Due to that, this side is named as the hadronic or physical side of the calculation. The calculation following the second path results in terms containing QCD degrees of freedom such as masses of the quarks, quark-gluon condensates, and QCD coupling constant, hence called as QCD side of the calculation. These two sides are matched via dispersion relation and give us the QCD sum rules through which we get the physical quantities under question. On both sides, there appear various Lorentz structures. In the match of the results, one considers one of the appropriate Lorentz structures present on each side, and to further improve the results, Borel transformation is applied to both sides to suppress the contributions of higher states and continuum.

In the calculation of the physical side, a complete set of intermediate states, with the same quantum numbers of the considered one, is substituted into the correlation function treating the interpolating current as annihilation and creation operator for the hadrons which have the same quantum numbers and quark contents of the interpolating current. Performing the integration over four-$x$, we obtain the correlator in this part as 
\begin{eqnarray}
\Pi^{\mathrm{Had}}(q)= \frac{\langle 0|\eta_{P_{css}}|P_{css}(q,s)\rangle \langle P_{css}(q,s)|\bar{\eta}_{P_{css}}|0\rangle}{m_{P_{css}}^2-q^2}+\cdots,
\label{eq:masshadronicside1}
\end{eqnarray}
where $\cdots$ is used to represent the higher states and continuum. $|P_{css}(q,s)\rangle$ represents the one-particle pentaquark state with momentum $q$ and spin $s$. The matrix elements in the Eq.~(\ref{eq:masshadronicside1}) have the following form given in terms of the current coupling constant $\lambda$ and Dirac spinor, $u_{P_{css}}(q,s)$, with spin $s$:
\begin{eqnarray}
\langle 0|\eta_{P_{css}}|P_{css}(q,s)\rangle &=& \lambda u_{P_{css}}(q,s).
\label{eq:matrixelement1}
\end{eqnarray}
When the last equation is substituted inside Eq.~(\ref{eq:masshadronicside1}), with the summation over the spin
\begin{eqnarray}
\sum_s u_{P_{css}}(q,s)\bar{u}_{P_{css}}(q,s)=\not\!q+m_{P_{css}},
\end{eqnarray}
the result of the hadronic side turns into
\begin{eqnarray}
\Pi^{\mathrm{Had}}(q)=\frac{\lambda^2(\not\!q+m_{P_{css}})}{m_{P_{css}}^2-q^2}+\cdots .
\end{eqnarray}
The hadronic side finally takes the following form after the Borel transformation:
\begin{eqnarray}
\tilde{\Pi}^{\mathrm{Had}}(q)=\lambda^2e^{-\frac{m_{P_{css}}^2}{M^2}}(\not\!q+m_{P_{css}})+\cdots,
\end{eqnarray}
where we represent the Borel transformed form of the correlator by $\tilde{\Pi}^{\mathrm{Had}}(q)$ and the Borel mass by $M^2$. In the final result we have two Lorentz structures, $\not\!q$ and $I$ that can be considered in the analyses. We take into account both of these structures and give the average value of the results obtained from each.

What comes next is the calculation of the correlator for the QCD side. For the calculation of this side, the interpolating current is used explicitly inside the Eq.~(\ref{eq:CorrFmassPc}), and Wick's theorem is applied to make the possible contractions between the quark fields. This operation turns the correlator into the one written in terms of the light and heavy quark propagators, which reads as
\begin{eqnarray}
\Pi^{\mathrm{QCD}}(q)&=&i\int d^4x e^{iqx} \epsilon_{ila}\epsilon_{ijk}\epsilon_{lmn}\epsilon_{i'l'a'}\epsilon_{i'j'k'}\epsilon_{l'm'n'}\Big\{-\mathrm{Tr}[S_s(x)^{kk'}\gamma_{5}CS_u^{T}{}^{jj'}(x)C\gamma_{5}]\mathrm{Tr}[S_c(x)^{nn'}\gamma_{5}CS_s^{T}{}^{mm'}(x)C\gamma_{5}]\nonumber\\
&+&
\mathrm{Tr}[S_s(x)^{mk'}\gamma_{5}CS_u^{T}{}^{jj'}(x)C\gamma_{5} S_s(x)^{km'}\gamma_{5}CS_c^{T}{}^{nn'}(x)C\gamma_{5}]\Big\}  C S_{c}^{T}{}^{aa'}(-x)C.
\label{Eq:PiQCDmass}
\end{eqnarray}   
For the calculation of this side, we use the light and heavy quark propagators explicitly in Eq.~(\ref{Eq:PiQCDmass}) as follows:
\begin{eqnarray}
S_{q,}{}_{ab}(x)&=&i\delta _{ab}\frac{\slashed x}{2\pi ^{2}x^{4}}-\delta _{ab}%
\frac{m_{q}}{4\pi ^{2}x^{2}}-\delta _{ab}\frac{\langle \overline{q}q\rangle
}{12} +i\delta _{ab}\frac{\slashed xm_{q}\langle \overline{q}q\rangle }{48}%
-\delta _{ab}\frac{x^{2}}{192}\langle \overline{q}g_{\mathrm{s}}\sigma
Gq\rangle +i\delta _{ab}\frac{x^{2}\slashed xm_{q}}{1152}\langle \overline{q}%
g_{\mathrm{s}}\sigma Gq\rangle  \notag \\
&&-i\frac{g_{\mathrm{s}}G_{ab}^{\alpha \beta }}{32\pi ^{2}x^{2}}\left[ %
\slashed x{\sigma _{\alpha \beta }+\sigma _{\alpha \beta }}\slashed x\right]
-i\delta _{ab}\frac{x^{2}\slashed xg_{\mathrm{s}}^{2}\langle \overline{q}%
q\rangle ^{2}}{7776} ,  \label{Eq:qprop}
\end{eqnarray}%
and
\begin{eqnarray}
S_{c,{ab}}(x)&=&\frac{i}{(2\pi)^4}\int d^4k e^{-ik \cdot x} \left\{
\frac{\delta_{ab}}{\!\not\!{k}-m_Q}
-\frac{g_sG^{\alpha\beta}_{ab}}{4}\frac{\sigma_{\alpha\beta}(\!\not\!{k}+m_Q)+
(\!\not\!{k}+m_Q)\sigma_{\alpha\beta}}{(k^2-m_Q^2)^2}\right.\nonumber\\
&&\left.+\frac{\pi^2}{3} \langle \frac{\alpha_sGG}{\pi}\rangle
\delta_{ij}m_Q \frac{k^2+m_Q\!\not\!{k}}{(k^2-m_Q^2)^4}
+\cdots\right\},
 \label{Eq:Qprop}
\end{eqnarray}
where $G^{\alpha\beta}_{ab}=G^{\alpha\beta}_{A}t_{ab}^{A}$ and $GG=G_{A}^{\alpha\beta}G_{A}^{\alpha\beta}$ with $a,~b=1,2,3$ and $A=1,2,\cdots,8$ and $t^A=\frac{\lambda^A}{2}$ where $\lambda^A$ is the  Gell-Mann matrices.The sub-index $q$ is used to represent $u$ ($s$) quark. After the application of the Fourier and Borel transformations we get 
\begin{eqnarray}
\tilde{\Pi}_i^{\mathrm{QCD}}(s_0,M^2)=\int_{(2m_c+2m_s)^2}^{s_0}dse^{-\frac{s}{M^2}}\rho_i(s)+\Gamma_i(M^2),
\label{Eq:Cor:QCD1}
\end{eqnarray}
where $\rho_i(s)$ are the spectral densities. We obtain the spectral densities from the imaginary parts of the results as $\frac{1}{\pi}\mathrm{Im}\Pi_i^{\mathrm{QCD}}$ with $i$ representing Lorentz structures, $\not\!q$ or $I$. In order not to overwhelm the article with long expressions we will not present the explicit forms of $\rho_i(s)$ and $\Gamma_i(M^2)$ here and focus on  just the results obtained from them. In Eq.~(\ref{Eq:Cor:QCD1}), $s_0$ represents the threshold parameter that arises after the application of the continuum subtraction using the quark-hadron duality assumption, and $M^2$ represents the Borel parameter. To obtain the physical quantities, mass and the current coupling constant, we match the results obtained from both the hadronic  and QCD sides considering the coefficient of the same Lorentz structure. The results for both structures are:
\begin{eqnarray}
\lambda^2e^{-\frac{m_{P_{css}}^2}{M^2}}&=&\tilde{\Pi}_{\not\!q}^{\mathrm{QCD}}(s_0,M^2),
\label{sumrulmasseq}
\end{eqnarray}
and
\begin{eqnarray}
\lambda^2m_{P_{css}}e^{-\frac{m_{P_{css}}^2}{M^2}}&=&\tilde{\Pi}_{I}^{\mathrm{QCD}}(s_0,M^2).
\label{sumrulemassI}
\end{eqnarray}
In principle, for the analyses, any of the Lorentz structure can be used. In our calculation we extract the results from both of them and obtain the final quantities by considering their averages. To obtain the results we need to use some input parameters which are presented in Table~\ref{tab:Inputs}.
\begin{table}[h!]
\begin{tabular}{|c|c|}
\hline\hline
Parameters & Values \\ \hline\hline
$m_{c}$                                     & $1.27\pm 0.02~\mathrm{GeV}$ \cite{Zyla:2020zbs}\\
$m_{b}$                                     & $4.18^{+0.03}_{-0.02}~\mathrm{GeV}$ \cite{Zyla:2020zbs}\\
$m_{s}$                                     & $93^{+11}_{-5}~\mathrm{MeV}$ \cite{Zyla:2020zbs}\\
$\langle \bar{q}q \rangle (1\mbox{GeV})$    & $(-0.24\pm 0.01)^3$ $\mathrm{GeV}^3$ \cite{Belyaev:1982sa}  \\
$\langle \bar{s}s \rangle $                 & $0.8\langle \bar{q}q \rangle$ \cite{Belyaev:1982sa} \\
$m_{0}^2 $                                  & $(0.8\pm0.1)$ $\mathrm{GeV}^2$ \cite{Belyaev:1982sa}\\
$\langle \overline{q}g_s\sigma Gq\rangle$   & $m_{0}^2\langle \bar{q}q \rangle$ \\
$\langle \frac{\alpha_s}{\pi} G^2 \rangle $ & $(0.012\pm0.004)$ $~\mathrm{GeV}^4 $\cite{Belyaev:1982cd}\\
$m_{J/\psi}$                                & $(3096.900\pm0.006)~\mathrm{MeV}$ \cite{Zyla:2020zbs}\\
$m_{D_s^-}$                                 & $(1968.35\pm 0.07)~\mathrm{MeV}$ \cite{Zyla:2020zbs}\\
$m_{\Xi^0}$                                 & $( 1314.86\pm 0.20 )~\mathrm{MeV}$ \cite{Zyla:2020zbs}\\
$m_{\Xi_c^+}$                               & $( 2467.71\pm 0.23 )~\mathrm{MeV}$ \cite{Zyla:2020zbs}\\
$\lambda_{\Xi^0}$                           & $(3.8 \pm 0.2)\times 10^{-2}~\mathrm{GeV}^3$ \cite{Wang:2007yt}\\
$\lambda_{\Xi_c^+}$                         & $(0.027 \pm 0.008)~\mathrm{GeV}^3$ \cite{Wang:2010fq}\\
$f_{J/\psi}$                                & $(481\pm36)~\mathrm{MeV}$ \cite{Veliev:2011kq}\\
$f_{D_s^-}$                                   & $(240\pm10)~\mathrm{MeV}$ \cite{Wang:2015mxa}\\
\hline\hline
\end{tabular}%
\caption{Some input parameters used in the analyses of mass and coupling constants.}
\label{tab:Inputs}
\end{table} 
This table also contains the parameters necessary for the coupling constant calculations of the next section. Besides the input parameters given in Table~\ref{tab:Inputs}, the mass calculations contain two more auxiliary parameters, which are the threshold parameter $s_0$ and the Borel parameter $M^2$. The proper intervals of these parameters are determined from the analyses based on the standard criteria of the sum rule method. One of the criteria is a mild dependence of the results on these parameters. The second one is the dominance of the contribution of the considered state over the higher ones and the continuum. And the last one is the convergence of the operator product expansion (OPE). To obtain the proper intervals, we consider these criteria and get the regions for which these criteria are satisfied. The working interval of the Borel parameter $M^2$ can be deduced by imposing the higher-order terms in the OPE side to have small contributions, and the dominant contribution comes from the lowest state. We analyze the results accordingly and deduce the interval of the Borel parameters as:
\begin{eqnarray}
5.0~\mbox{GeV}^2\leq M^2\leq 7.0~\mbox{GeV}^2,
\end{eqnarray}
and we extract the interval for the threshold parameter, which also has a connection to the energy of the possible excited state, as:
\begin{eqnarray}
&24.5~\mbox{GeV}^2 \leq s_0 \leq 26.5~\mbox{GeV}^2.&
\end{eqnarray} 
With the application of these parameters and the input parameters given in Table~\ref{tab:Inputs}, finally, we get the mass and current coupling constant as
\begin{eqnarray}
m_{P_{css}}=4600\pm 155~\mathrm{MeV}~~~~~~\mathrm{and}~~~~~~\lambda_{P_{css}}=(0.81\pm 0.21)\times10^{-3}~\mathrm{GeV}^6.
\end{eqnarray}
The results obtained here serve as input parameters for the next sections in which we discuss the strong decays of the considered state to  $\Xi^0 J/\psi$ and $\Xi_c^+D^-$  final states.

\section{The strong decay $P_{css} \rightarrow \Xi^0 J/\psi $}\label{III}

This section is devoted to the calculation of the width for the decay $P_{css}\rightarrow \Xi^0 J/\psi$. The calculation is performed via the three-point correlation function which has the following form:
\begin{equation}
\Pi_{\mu} (p, q)=i^2\int d^{4}xe^{-ip\cdot
x}\int d^{4}ye^{ip'\cdot y}\langle 0|\mathcal{T} \{\eta^{\Xi^0}(y)
\eta_{\mu}^{J/\psi}(0)\bar{\eta}^{P_{css}}(x)\}|0\rangle.
\label{eq:CorrF1Pc}
\end{equation}
In this three-point correlation function, $\eta^{P_{css}}$ still represents the interpolating current for the candidate $P_{css}$ state, which is given in Eq.~(\ref{CurrentPcss}). The interpolating currents of final states, $\Xi^0$ and $J/\psi$ states, are represented by $\eta^{\Xi^0}$ and $\eta_{\mu}^{J/\psi}$, respectively, and these currents have the following forms
\begin{eqnarray}
\eta^{\Xi^0}&=&-2\epsilon^{lmn}\sum_{i=1}^{2}\Big[(s^{T}_l CA_1^i u_m)A_2^i s_n\Big],\nonumber\\
\eta_{\mu}^{J/\psi}&=&\bar{c}_l\gamma_{\mu}c_l,
\label{InterpFields}
\end{eqnarray}
with the sub-indices, $l,~m,~n$, we represent the color indices, and $u,~s,~c$ are the quark fields, $A_1^1=I$, $A_1^2=A_2^1=\gamma_5$, and $A_2^2=\beta$ is an arbitrary parameter, and $C$ is the charge conjugation operator. The parameter $\beta$ is fixed from the analyses of the results, and its working interval is to be presented. The standard steps of the QCD sum rule are followed again, and the physical quantities, which are the coupling constants $g_1$ and $g_2$ for the strong vertex, are obtained via the match of the coefficients of the same Lorentz structures attained from the hadronic and QCD sides.

According to the standard QCD sum rule procedure, we first calculate the hadronic side via insertion of the complete sets of hadronic states with the same quantum numbers of the interpolating fields, and the four integrals are performed to obtain the following representation
\begin{eqnarray}
\Pi _{\mu }^{\mathrm{Had}}(p, q)=\frac{\langle 0|\eta^{\Xi^0 }|\Xi^0(p',s')\rangle \langle 0|\eta_{\mu }^{J/\psi}|J/\psi(q)\rangle \langle J/\psi(q) \Xi^0(p',s')|P_{css}(p,s)\rangle \langle P_{css}(p,s)|\bar{\eta}^{P_{css}}|0\rangle }{(m_{\Xi^0}^2-p'^2)(m_{J/\psi}^2-q^2)(m_{P_{css}}^2-p^2)}+\cdots,  \label{eq:CorrF1Phys}
\end{eqnarray}
where $\cdots$ represents the contributions coming from the higher states and continuum, the $p$, $p'$, and $q$ are the momenta of the $P_{css}$ and $\Xi^0$ and $J/\psi$ states, respectively. To proceed, the matrix elements in the last result are required, and they are represented in terms of the masses and current coupling constants as:
\begin{eqnarray}
\langle 0|\eta^{P_{css}}|P_{css}(p,s)\rangle &=&\lambda_{P_{css}} u_{P_{css}}(p,s),
\nonumber\\
\langle 0|\eta^{\Xi^0 }|\Xi(p',s')\rangle &=&\lambda_{\Xi^0} u_{\Xi^0}(p',s'),
\nonumber\\
\langle 0|\eta_{\mu }^{J/\psi}|J/\psi(q)\rangle &=&f_{J/\psi} m_{J/\psi} \varepsilon_{\mu}.
\label{eq:ResPcss}
\end{eqnarray}
The $ \varepsilon_{\mu} $, $ f_{J/\psi} $, $\lambda_{P_{css}}$, $\lambda_{\Xi^0}$ in the matrix elements correspond to the polarization vector and the decay constant of the $J/\psi$ state and the current coupling constants of the $P_{css}$ and $\Xi^0$ states, respectively. $|P_{css}(p,s)\rangle$ represents the one-particle pentaquark state whose spinor is given by $ u_{P_{css}}$ and the spinor for the $\Xi^0$ state is $ u_{\Xi^0}$. The matrix element defining the coupling constants, $g_1$ and $g_2$, is
\begin{eqnarray}
\langle J/\psi(q) \Xi^0(p',s')|P_{css}(p,s)\rangle = \epsilon^{* \nu}\bar{u}_{\Xi^0}(p',s')\big[g_1\gamma_{\nu}-\frac{i\sigma_{\nu\alpha}}{m_{\Xi^0}+m_{P_{css}}}q^{\alpha}g_2\big]\gamma_5 u_{P_{css}}(p,s).
\label{eq:coupling}
\end{eqnarray}
When we substitute the matrix elements given in Eq.~(\ref{eq:ResPcss}) and Eq.~(\ref{eq:coupling}) into the Eq.~(\ref{eq:CorrF1Phys}) together with the following summation over the spins of the spinors and polarization vector
\begin{eqnarray}
\sum_{s}u_{P_{css}}(p,s)\bar{u}_{P_{css}}(p,s)&=&({\slashed
p}+m_{P_{css}}),\nonumber \\
\sum_{s'}u_{\Xi^0}(p',s')\bar{u}_{\Xi^0}(p',s')&=&({\slashed
p'}+m_{\Xi^0}), \nonumber\\
\varepsilon_{\alpha}\varepsilon^*_{\beta}&=&-g_{\alpha\beta}+\frac{q_\alpha q_\beta}{m_{J/\psi}^2},
\label{eq:SumPc}
\end{eqnarray}
we get the physical side as
\begin{eqnarray}
\tilde{\Pi}_{\mu }^{\mathrm{Had}}(p, q)&=&e^{-\frac{m_{P_{css}^2}}{M^2}}e^{-\frac{m_{\Xi^0}^2}{M'^2}}\frac{f_{J/\psi} \lambda_{\Xi^0} \lambda_{P_{css}}  m_{\Xi^0} }{ m_{J/\psi} (m_{\Xi^0} + m_{P_{css}}) (m_{J/\psi}^2 + Q^2)}\big[ g_1 (m_{\Xi^0} + m_{P_{css}})^2+g_2 m_{J/\psi}^2 \big] \not\!p p'_{\mu} \gamma_5
 \nonumber\\
&+&e^{-\frac{m_{P_{css}^2}}{M^2}}e^{-\frac{m_{\Xi^0}^2}{M'^2}}\frac{f_{J/\psi} \lambda_{\Xi^0} \lambda_{P_{css}}  }{ m_{J/\psi} (m_{\Xi^0} + m_{P_{css}}) (m_{J/\psi}^2 + Q^2)}\big[ -g_1 (m_{\Xi^0} + m_{P_{css}})\big( m_{P_{css}}^3+ m_{P_{css}} Q^2\nonumber\\
&+& m_{\Xi^0}(2  m_{J/\psi}^2+ m_{\Xi^0}^2+Q^2)\big)+g_2 m_{J/\psi}^2(m_{P_{css}}^2+m_{P_{css}} m_{\Xi^0}-m_{\Xi^0}^2+Q^2) \big]  p_{\mu} \gamma_5 \nonumber\\
&+&
\mathrm{other~structures}+\cdots.
\label{eq:had}
\end{eqnarray}
In Eq.~(\ref{eq:had}), $Q^2=-q^2$, $M^2$ and $M'^2$ are the Borel parameters which are determined from the analysis of the results following standard criteria of the QCD sum rule method. In this part, only the terms containing the Lorentz structures $\not\!p p'_{\mu} \gamma_5$, $p_{\mu} \gamma_5 $ are given explicitly, and the result contains more Lorentz structures than the given ones. We present only these because we apply them directly in our further analyses. The contributions coming from the other structures and excited states and continuum are represented as $\mathrm{other~structures}+\cdots$.

After the physical side, the QCD side is calculated using the interpolating fields given in Eqs.~(\ref{CurrentPcss}) and (\ref{InterpFields}) explicitly in the correlation function, Eq.~(\ref{eq:CorrF1Pc}). Performing the possible contractions between the quark fields via Wick's theorem, the result is achieved in terms of the light and heavy quark propagators. This part results in
\begin{eqnarray}
\Pi_{\mu }^\mathrm{OPE}(p,p',q)&=&i^2\int d^{4}xe^{-ip\cdot x}\int d^{4}ye^{ip'\cdot y}\epsilon^{klm}\epsilon^{i'l'a'}\epsilon^{i'j'k'}\epsilon^{l'm'n'} 2\sum_{i=1}^2\bigg\{Tr[\gamma_{5}CS_{u}^{Tkj'}(y-x)(C A_1^i)^T S_{s}^{lk'}(y-x)]\nonumber\\
&\times& A_2^i S_s^{mm'}(y-x) \gamma_{5}C S_{c}^{Tnn'}(-x) C \gamma_{\mu} C S_{c}^{Ta'n}(x)C - A_2^iS_{s}^{mk'}(y-x)\gamma^{5}CS_{u}^{Tkj'}(y-x) (C A_1^i)^T S_{s}^{lm'}(y-x)\nonumber\\
&\times& 
\gamma_{5}C S_{c}^{Tnn'}(-x) C \gamma_{\mu}C S_{c}^{Ta'n}(x)C\bigg\}.\label{eq:CorrF1Theore}
\end{eqnarray}
In the last equation $S_q^{ab}(x)=S_{u(s)}^{ab}(x)$ and $S_{c}^{ab}(x)$ are the light and heavy quark propagators given in the Eqs.~(\ref{Eq:qprop}) and ~(\ref{Eq:Qprop}), respectively. This side emerges with the same Lorentz structures as in the hadronic side. We again isolate the coefficients of the structures used in the analyses, which are $\not\!p p'_{\mu} \gamma_5$, $p_{\mu} \gamma_5$,  and represent the QCD side as 
\begin{eqnarray}
\Pi_{\mu }^{OPE}(p,q)&=&\Pi_1\,\not\!p p'_{\mu} \gamma_5 +
\Pi_2\, p_{\mu} \gamma_5 +\mathrm{other \,\,\, structures}, \label{eq:PiOPE}
\end{eqnarray}
where the other structures term represents the contributions coming from the other Lorentz structures. The calculations of the coefficients $\Pi_i$ are performed using the light and the heavy quark propagators explicitly in the Eq.~(\ref{eq:CorrF1Theore}). Then the results are transformed to momentum space, and the four integrals are computed. Finally, after the application of the Borel transformation, the spectral densities are obtained from the imaginary parts of the results as $\rho_i(s,s',q^2)=\frac{1}{\pi}Im[\Pi_i]$ and used in the following dispersion relation:
\begin{eqnarray}
\Pi_{i}=\int ds\int
ds'\frac{\rho_{i}^{\mathrm{pert}}(s,s',q^{2})+\rho_{i}^{\mathrm{non-pert}}(s,s',q^{2})}{(s-p^{2})
(s'-p'^{2})}. \label{eq:Pispect}
\end{eqnarray}
The last equation represents the final form of the QCD side of the calculation with $ \rho_{i}^{\mathrm{pert}}(s,s',q^{2}) $ and $ \rho_{i}^{\mathrm{non-pert}}(s,s',q^{2}) $ being the perturbative and non-perturbative parts of the spectral densities, respectively, where $i=1,2,..,12$. Again we will not give the lengthy mathematical expressions of the results here and focus on their analyses.

After the computation of each side, the matches of the results coming from the coefficients of the same Lorentz structures are used to attain the physical quantities, $g_1$ and $g_2$. From the considered structures, we obtained the following two coupled relations:
\begin{eqnarray}
&&e^{-\frac{m_{P_{css}^2}}{M^2}}e^{-\frac{m_{\Xi^0}^2}{M'^2}}\frac{f_{J/\psi} \lambda_{\Xi^0} \lambda_{P_{css}}  m_{\Xi^0} }{ m_{J/\psi} (m_{\Xi^0} + m_{P_{css}}) (m_{J/\psi}^2 + Q^2)}\big[ g_1 (m_{\Xi^0} + m_{P_{css}})^2+g_2 m_{J/\psi}^2 \big] =
\tilde{\Pi}_1,
\\
&&e^{-\frac{m_{P_{css}^2}}{M^2}}e^{-\frac{m_{\Xi^0}^2}{M'^2}}\frac{f_{J/\psi} \lambda_{\Xi^0} \lambda_{P_{css}}  }{ m_{J/\psi} (m_{\Xi^0} + m_{P_{css}}) (m_{J/\psi}^2 + Q^2)}
\big[ -g_1 (m_{\Xi^0} + m_{P_{css}})\big( m_{P_{css}}^3
+ m_{P_{css}} Q^2 m_{\Xi^0}(2  m_{J/\psi}^2+ m_{\Xi^0}^2+Q^2)\big)\nonumber\\
&+&g_2 m_{J/\psi}^2(m_{P_{css}}^2+m_{P_{css}} m_{\Xi^0}-m_{\Xi^0}^2+Q^2) \big] =
\tilde{\Pi}_2,
\label{eq:SR}
\end{eqnarray}
where $\tilde{\Pi}_i$ is used to represent the $\Pi_i$ after the Borel transformations. The solutions of these two relations for the $g_1$ and $g_2$ have the following forms:
\begin{eqnarray}
g_1&=&
e^{\frac{m_{P_{css}}^2}{M^2}}e^{\frac{m_{\Xi^0}^2}{M'^2}}\frac{m_{J/\psi}(m_{J/\psi}^2+Q^2)
\left[(m_{P_{css}}^2+m_{P_{css}}m_{\Xi^0}-m_{\Xi^0}^2+Q^2)\tilde{\Pi}_1 -m_{\Xi^0}\tilde{\Pi}_2 \right]}
{2f_{J/\psi}\lambda_{\Xi^0}\lambda_{P_{css}}m_{\Xi^0}[m_{P_{css}}^3+m_{P_{css}}^2m_{\Xi^0}+m_{P_{css}}Q^2+m_{\Xi^0}(m_{J/\psi}^2+Q^2)]},\nonumber \\
g_2&=& e^{\frac{m_{P_{css}}^2}{M^2}}e^{\frac{m_{\Xi^0}^2}{M'^2}}(m_{P_{css}}+m_{\Xi^0})(m_{J/\psi}^2+Q^2)\nonumber\\
&\times&\frac{
\left[[m_{P_{css}}^3+m_{P_{css}}Q^2+m_{\Xi^0}(2m_{J/\psi}^2+m_{\Xi^0}^2+Q^2)]\tilde{\Pi}_1+m_{\Xi^0}(m_{P_{css}}+m_{\Xi^0})\tilde{\Pi}_2  \right]}
{2f_{J/\psi}\lambda_{\Xi^0}\lambda_{P_{css}}m_{\Xi^0}m_{J/\psi}[m_{P_{css}}^3+m_{P_{css}}^2m_{\Xi^0}+m_{P_{css}} Q^2+m_{\Xi^0}(m_{J/\psi}^2+Q^2)]}.
  \label{eq:g1}
\end{eqnarray}
These results also require some input parameters, which are given in Table~\ref{tab:Inputs} and there enters five more auxiliary parameters into the coupling constant calculations, which are Borel parameters $M^2$ and $M'^2$, threshold parameters $s_0$ and $s'_0$ and the arbitrary parameter $\beta$ present in the interpolating field of the $\Xi^0$ baryon. To determine their working intervals, we seek their proper regions satisfying the standard QCD sum rule criteria whose details are also stated in the mass calculation part of this work. We extract the regions for the auxiliary parameters from our analyses as
\begin{eqnarray}
5.0\ \mathrm{GeV}^{2}\leq M^{2}& \leq & 7.0\ \mathrm{GeV}^{2},
\nonumber \\
1.6\ \mathrm{GeV}^{2}\leq M'^{2}& \leq & 3.0\ \mathrm{GeV}^{2},
\label{Eq:MsqMpsq}
\end{eqnarray}%
and
\begin{eqnarray}
24.5\,\,\mathrm{GeV}^{2}&\leq& s_{0}\leq 26.5\,\,\mathrm{GeV}^{2},
\nonumber \\
2.0\,\,\mathrm{GeV}^{2}&\leq& s'_{0}\leq 2.5\,\,\mathrm{GeV}^{2}.
\label{Eq:s0s0p}
\end{eqnarray}
As for the parameter $\beta$, its interval is fixed by a parametric plot analysis of the results. In this analysis, the results are plotted as a function of $\cos\theta$, defining $\beta=\tan \theta$, and the regions with least sensitivity to the variation of $\cos\theta$ are selected, which are obtained as
\begin{eqnarray}
-1.0\leq\cos\theta\leq -0.5 ~~~~~\mbox{and} ~~~~~~0.5\leq \cos\theta\leq 1.0 
\end{eqnarray}  
We use all these auxiliary parameters and input parameters given in Table~\ref{tab:Inputs} to analyze our results and attain the coupling constants for the decay channel under question.  The results obtained from these sum rule analyses are reliable in a range of $Q^2$ values. Therefore, to expand it to the region of interest, we apply a proper fit function which has the following form
\begin{eqnarray}
g_i(Q^2)&=& g_{0}e^{c_1\frac{Q^2}{m_{P_{css}}^2}+c_2(\frac{Q^2}{m_{P_{css}}^2})^2}.
\end{eqnarray}
The values obtained for fit parameters $g_0$, $c_1$ and $c_2$ are given in the Table~\ref{tab:FitParam}.
\begin{table}[tbp]
\begin{tabular}{|c|c|c|c|c|c|}
\hline\hline
Decay Channel& Coupling Constant  &$ g_0$ & $c_1$ & $c_2$ & $g_i(-m_{J/\psi}^2) $ \\ \hline\hline
\multirow{2}{*}{ $P_{css} \rightarrow \Xi^0 J/\psi$}  &$g_1$&$12.80  $ &$1.57$ & $-0.91$ & $5.21 \pm 0.78$\\ \cline{2-6} 
& $g_2$&$7.80$ & $1.40$ & $4.02$ & $9.45 \pm 1.43$\\
\hline\hline 
Decay Channel&Coupling Constant  &$ g_0$ & $c_1$ & $c_2$ & $g(-m_{D_s^-}^2) $ \\ \hline\hline
$P_{css} \rightarrow \Xi_c^+ D_s^-$ &$g$&$0.92$ &$5.98$ & $-8.16$ & $0.23 \pm 0.03$\\
\hline\hline
\end{tabular}%
\caption{ Values of the parameters for the fit functions of
coupling constants, $g_1$ and $g_2$ for $P_{css} \rightarrow \Xi^0 J/\psi$ decay and and $g$ for $P_{css} \rightarrow \Xi_c^+ D_s^-$ decay and their values obtained from the fit functions at $Q^2=-m_{J/\psi}^2$ and $Q^2=-m_{D_s}^2$, respectively.} \label{tab:FitParam}
\end{table}

As a final task, the coupling constants are applied to get the width of the considered decay channel, $P_{css} \rightarrow \Xi^0 J/\psi $ with the following width formula
\begin{eqnarray}
\Gamma &=& \frac{f(m_{P_{css}},m_{J/\psi},m_{\Xi^0}) }{16\pi m_{P_{css}}^2}\Bigg[-\frac{2 (m_{J/\psi}^2 - (m_{\Xi^0} + m_{P_{css}})^2)}{m_{J/\psi}^2 (m_{\Xi^0} + m_{P_{css}})^2}\Big(g_2^2 m_{J/\psi}^2 (m_{J/\psi}^2 + 2 (m_{\Xi^0} - m_{P_{css}})^2) 
\nonumber\\
&+& 
6 g_1 g_2 m_{J/\psi}^2 (m_{\Xi^0} - m_{P_{css}}) (m_{\Xi^0} + m_{P_{css}}) + 
g_1^2 (2 m_{J/\psi}^2 + (m_{\Xi^0} - m_{P_{css}})^2) (m_{\Xi^0} + m_{P_{css}})^2 \Big)\Bigg],
\label{Eq:DW}
\end{eqnarray}
where
\begin{eqnarray}
f(x,y,z)&=&\frac{1}{2x}\sqrt{x^4+y^4+z^4-2x^2y^2-2x^2z^2-2y^2z^2}.\label{functionf}
\end{eqnarray} 
The corresponding width is obtained as
\begin{eqnarray}
\Gamma(P_{css} \rightarrow \Xi^0 J/\psi )&=& 9.52\pm 2.85~\mathrm{MeV}.
 \label{Eq:DWNegativeParity}
\end{eqnarray}

\section{The strong decay $P_{css} \rightarrow \Xi_c^+ D_s^- $}\label{IV}
To see the effects of the other channels on the total width of the possible double strange pentaquark state, we also consider its decay channel $P_{css} \rightarrow \Xi_c^+ D_s^- $ in our calculation. This analysis may shed light on the identification of the possible main decay mode. Considering the available phase space, we take into account $P_{css} \rightarrow \Xi_c^+ D_s^- $ decay and follow the similar calculation strategy of the previous section to obtain the corresponding decay width. For the three-point correlator for this decay, we again apply the one given in Eq.~(\ref{eq:CorrF1Pc}), and we replace the $\Pi _{\mu }(p, q)$ with $\Pi(p, q)$. And also, instead of the interpolating fields  $\eta_{\mu }^{J/\psi} $ and $\eta^{\Xi^0 }$ present in the Eq.~(\ref{eq:CorrF1Pc}), we use the relevant interpolating fields  $\eta^{\Xi_c^+ }$ and $\eta^{D_s^-}$  with the following explicit forms:
\begin{eqnarray}
\eta^{\Xi_c^+}&=&\epsilon^{abc}u^{T}_a C \gamma_5 s_bc_c,\nonumber\\
\eta^{D_s^-}&=&\bar{c}_i\gamma_{5}s_i,
\label{InterpFields2}
\end{eqnarray}
where the sub-indices, $a,~b,~c,~i$ are the color indices and $u,~s,~c$ are the quark fields. The coupling constant $g$ for this transition is calculated following the similar steps of the previous section. The calculation of the hadronic side results in
\begin{eqnarray}
\Pi^{\mathrm{Had}}(p, q)=\frac{\langle 0|\eta^{\Xi_c^+ }|\Xi_c(p',s')\rangle \langle 0|\eta^{D_s^-}|D_s(q)\rangle \langle D_s(q) \Xi_c(p',s')|P_{css}(p,s)\rangle \langle P_{css}(p,s)|\bar{\eta}^{P_{css}}|0\rangle }{(m_{\Xi_c^+}^2-p'^2)(m_{D_s^-}^2-q^2)(m_{P_{css}}^2-p^2)}+\cdots,  \label{eq:CorrF1Phys2}
\end{eqnarray}
with the matrix elements
\begin{eqnarray}
\langle 0|\eta^{\Xi_c^+}|\Xi_c(p',s')\rangle &=&\lambda_{\Xi_c^+} u_{\Xi_c^+}(p',s'),
\nonumber\\
\langle 0|\eta^{D_s^-}|D_s(q)\rangle &=&\frac{f_{D_s^-} m_{D_s^-}^2}{m_s+m_c},
\label{eq:ResPcss2}
\end{eqnarray}
where  $f_{D_s^-} $ and $\lambda_{\Xi_c^+}$ in these matrix elements are the decay constant of the $D_s^-$ meson and residue of the $\Xi_c^+$ state, respectively.  To define the coupling constant $g$ of the decay we use 
\begin{eqnarray}
\langle D_s(q) \Xi_c(p',s')|P_{css}(p,s)\rangle = g \bar{u}_{\Xi_c^+}(p',s') u_{P_{css}}(p,s).
\label{eq:coupling2}
\end{eqnarray} 
Substitution of these matrix elements into the Eq.~(\ref{eq:CorrF1Phys2}) and using the summation over the spins of the spinors using Eq.~({\ref{eq:SumPc}}) just by replacement of $\Xi^0$ by $\Xi_c^+$ gives the physical side as
\begin{eqnarray}
\tilde{\Pi}^{\mathrm{Had}}(p, q)&=&e^{-\frac{m_{P_{css}^2}}{M^2}}e^{-\frac{m_{\Xi_c^+}^2}{M'^2}}\frac{g f_{D_s^-}m_{D_s^-}^2 \lambda_{\Xi_c^+} \lambda_{P_{css}} }{(m_s+m_c) (m_{D_s^-}^2 + Q^2)} \not\!p\not\!p' +
\mathrm{other~structures}+\cdots.
\label{eq:had2}
\end{eqnarray}
The above result presents the term with the Lorentz structures $\not\!p\not\!p' $ that is used in the analysis of this decay, and the other terms and the contribution of the excited states are denoted by $\mathrm{other~structures}+\cdots$.

For the QCD side, that is obtained similarly as in the previous section,  the result is  
\begin{eqnarray}
\Pi^\mathrm{OPE}(p,p',q)&=&i^2\int d^{4}xe^{-ip\cdot x}\int d^{4}ye^{ip'\cdot y}\epsilon^{abc}\epsilon^{i'l'a'}\epsilon^{i'j'k'}\epsilon^{l'm'n'} \bigg\{Tr[S_{s}^{bk'}(y-x)\gamma_5C S_{u}^{Taj'}(y-x)C\gamma_{5}]\nonumber\\
&\times& S_c^{cn'}(y-x) \gamma_{5}C S_{s}^{Tim'}(-x) C \gamma_{5} C S_{c}^{Ta'i}(x)C -S_{c}^{cn'}(y-x)\gamma^{5}CS_{s}^{Tbm'}(y-x) C \gamma_5 S_{u}^{aj'}(y-x)\nonumber\\
&\times& 
\gamma_{5}C S_{s}^{Tik'}(-x) C \gamma_{5}C S_{c}^{Ta'i}(x)C\bigg\}.\label{eq:CorrF1Theore2}
\end{eqnarray}
The propagators present in this result are given in Eqs.~(\ref{Eq:qprop}) and (\ref{Eq:Qprop}), and the final form of the result contains the same Lorentz structures of the hadronic side. Taking the coefficient of the same Lorentz structure, $\not\!p'\not\!p$, from the OPE and hadronic sides and matching them, we get the corresponding coupling constant as:
\begin{eqnarray}
g&=&
e^{\frac{m_{P_{css}}^2}{M^2}}e^{\frac{m_{\Xi_c^+}^2}{M'^2}}\frac{(m_s+m_c)(m_{D_s^-}^2+Q^2)}
{f_{D_s^-}m_{D_s^-}^2 \lambda_{\Xi_c^+} \lambda_{P_{css}}}\tilde{\Pi}.
  \label{eq:g}
\end{eqnarray}
$\tilde{\Pi}$ is the result obtained for the QCD side after the Borel transformation. The input parameters necessary for this calculation are also given in Table~\ref{tab:Inputs}, and the auxiliary parameters, $M^2$  and $s_0$, are taken as in the previous section. The $M'^2$ is set from the analyses as:  
\begin{eqnarray}
\nonumber \\
3.0\ \mathrm{GeV}^{2}\leq M'^{2}& \leq & 4.0\ \mathrm{GeV}^{2},
\label{Eq:MsqMpsq2}
\end{eqnarray}%
and $s_0'$ is determined as
\begin{eqnarray}
7.6\,\,\mathrm{GeV}^{2}&\leq& s'_{0}\leq 8.4\,\,\mathrm{GeV}^{2},
\label{Eq:s0s0p2}
\end{eqnarray}
based on the standard criteria of the QCD sum rule method. 

The use of all the results with the mentioned input parameters leads to a coupling constant analysis that is again reliable in a region of $Q^2$ values, which necessitates a proper fit function as in the previous section. Here we apply the fit function with the same form as that of the Sec.~\ref{III}. The results for the fit parameters and the coupling constant obtained with the adopted fit function at $Q^2=-m_{D_s^-}^2$ for this decay mode are also presented in Table~\ref{tab:FitParam}.

For this decay channel, the width of the decay is obtained using the following formula in terms of the coupling constant $g$:
\begin{eqnarray}
\Gamma &=& \frac{
g^2 f(m_{P_{css}},m_{D_s^-},m_{\Xi_c^+}) }{8\pi m_{P_{css}}^2}[( m_{P_{css}} + m_{\Xi_c^+} )^2 - m_{D_s^-}^2)],
\label{Eq:DW2}
\end{eqnarray}
where the $f(m_{P_{css}},m_{D_s^-},m_{\Xi_c^+})$ is obtained from Eq.~(\ref{functionf}). The width for this channel is obtained as
\begin{eqnarray}
\Gamma(P_{css} \rightarrow \Xi_c^+ D_s^- )&=& 2.77\pm 0.72~\mathrm{MeV}.
 \label{Eq:DWvalue2}
\end{eqnarray}

As is seen from the present result, the width for the $P_{css} \rightarrow \Xi_c^+ D_s^- $ channel is very small compared to the $P_{css} \rightarrow \Xi^0 J/\psi$.


\section{Summary and conclusion}\label{Sum} 

The number of pentaquark candidate states is growing due to the improved analyses and the experimental techniques. Looking at the particle history, it is natural to expect such new pentaquark states with different quark substructures. Besides the state including a single strange quark $P_{cs}(4459)^0$ observed by the LHCb Collaboration~\cite{LHCb:2020jpq} in the $J/\psi \Lambda$ invariant mass distribution of the $\Xi_b^-\rightarrow J/\psi \Lambda K^-$ decay, the pentaquark states with double or triple strange quark content may exist and to be observed. By this expectation, in this work, we considered a possible pentaquark state with double strange quark. To analyze various properties of such a possible state, we applied the QCD sum rule method and used an interpolating current in the form of diquark-diquark-antiquark with spin-parity quantum numbers $J^P=\frac{1}{2}^-$. The QCD sum rule method was widely used in various analyses in literature and gave reliable predictions consistent with the experimental observations and the predictions of the other approaches.

In the first part, we calculated the mass for this state via two-point QCD sum rule. In the analyses we considered both of the present Lorentz structures and extracted average values for both the mass and current coupling constant of the state, which are $m_{P_{css}}=4600\pm 155~\mathrm{MeV}$ and $\lambda_{P_{css}}=(0.81\pm 0.21)\times 10^{-3}~\mathrm{GeV}^6$, respectively. Although the reported uncertainty for the  mass  well remains under the limits allowed by the method, it leaves a small  energy range  between the mass of the $  P_{css}$ and masses of the constituent particles considering all the uncertainties. 

The investigations in the literature for the presently observed pentaquark states indicate that the spectroscopic analyses may not be enough to fix the properties of such states. Therefore, we also expanded our analyses with the inclusion of the decay width computations for the strong decays of the considered state, $P_{css} \rightarrow \Xi^0 J/\psi$ and $P_{css} \rightarrow \Xi_c^+ D_s^-$. To obtain the corresponding widths, we need the coupling constants, $g_1$ and $g_2$ for former decay and $g$ for the later one, defining these transitions, and they were calculated using three-point QCD sum rule approach at $Q^2=-m_{J/\psi}^2$ and $Q^2=-m_{D_s^-}^2$, respectively. Then these results were applied to get the corresponding widths as $\Gamma(P_{css} \rightarrow \Xi^0 J/\psi)=9.52\pm 2.85~\mathrm{MeV}$ and $\Gamma(P_{css} \rightarrow  \Xi_c^+ D_s^-)=2.77\pm 0.72~\mathrm{MeV}$. 

For this pentaquark state one can also consider $P_{css} \rightarrow  \Xi'{}_c^+ D_s^-$ decay mode, however considering the masses of particles at final state, it is clear that the available phase space is smaller for this mode compared to the $P_{css} \rightarrow  \Xi_c^+ D_s^-$ channel. Due to this, the decay width for this mode is expected to be even  smaller. As a result, the total width is predicted to be $\Gamma=\Gamma(P_{css} \rightarrow \Xi^0 J/\psi)+\Gamma(P_{css} \rightarrow  \Xi_c^+ D_s^-)=12.29\pm 2.94~\mathrm{MeV}$ with $P_{css} \rightarrow \Xi^0 J/\psi$ being dominant decay mode. 

The result of mass and obtained  widths may shed light on the future observations of such states in $ J/\psi \Xi $ invariant mass. In the future, the experimentally observed such states may be compared with the obtained results to provide insights into the explanation of their sub-structures and possible quantum numbers.

\section*{ACKNOWLEDGEMENTS}
K. Azizi is thankful to Iran Science Elites Federation (Saramadan)
for the partial  financial support provided under the grant number ISEF/M/400150.




\begin{thebibliography}{99}

\bibitem{Gell-Mann} 
M. Gell-Mann, 
\href{https://www.sciencedirect.com/science/article/abs/pii/S0031916364920013?via%3Dihub}{ Phys. Lett. 8, 214 (1964)}.

\bibitem{Choi:2003ue} 
  S.~K.~Choi {\it et al.} [Belle Collaboration],
 \href{https://doi.org/10.1103/PhysRevLett.91.262001}{Phys.\ Rev.\ Lett.\  {\bf 91}, 262001 (2003)}
\href{https://arxiv.org/abs/hep-ex/0309032}{[hep-ex/0309032]}.


\bibitem{BaBar:2004oro}
B.~Aubert \textit{et al.} [BaBar],
\href{https://journals.aps.org/prd/abstract/10.1103/PhysRevD.71.071103}{Phys. Rev. D \textbf{71}, 071103 (2005)}
\href{https://arxiv.org/abs/hep-ex/0406022}{[arXiv:hep-ex/0406022 [hep-ex]]}.



\bibitem{CDF:2003cab}
D.~Acosta \textit{et al.} [CDF],
\href{https://journals.aps.org/prl/abstract/10.1103/PhysRevLett.93.072001}{Phys. Rev. Lett. \textbf{93}, 072001 (2004)}
\href{https://arxiv.org/abs/hep-ex/0312021}{[arXiv:hep-ex/0312021 [hep-ex]]}.


\bibitem{CDF:2009nxk}
T.~Aaltonen \textit{et al.} [CDF],
\href{https://journals.aps.org/prl/abstract/10.1103/PhysRevLett.103.152001}{Phys. Rev. Lett. \textbf{103}, 152001 (2009)}
\href{https://arxiv.org/abs/0906.5218}{[arXiv:0906.5218 [hep-ex]]}.


\bibitem{D0:2004zmu}
V.~M.~Abazov \textit{et al.} [D0],
\href{https://journals.aps.org/prl/abstract/10.1103/PhysRevLett.93.162002}{Phys. Rev. Lett. \textbf{93}, 162002 (2004)}
\href{https://arxiv.org/abs/hep-ex/0405004}[arXiv:hep-ex/0405004 [hep-ex]].



\bibitem{LHCb:2011zzp}
R.~Aaij \textit{et al.} [LHCb],
\href{https://link.springer.com/article/10.1140%2Fepjc%2Fs10052-012-1972-7}{Eur. Phys. J. C \textbf{72}, 1972 (2012)}
\href{https://arxiv.org/abs/1112.5310}{[arXiv:1112.5310 [hep-ex]]}.


\bibitem{CMS:2013fpt}
S.~Chatrchyan \textit{et al.} [CMS],
\href{https://link.springer.com/article/10.1007%2FJHEP04%282013%29154}{JHEP \textbf{04}, 154 (2013)}
\href{https://arxiv.org/abs/1302.3968}{[arXiv:1302.3968 [hep-ex]]}.


\bibitem{Aaij:2015tga} 
  R.~Aaij {\it et al.} [LHCb Collaboration],
\href{https://doi.org/10.1103/PhysRevLett.115.072001}{  Phys.\ Rev.\ Lett.\  {\bf 115}, 072001 (2015)}
\href{https://arxiv.org/abs/1507.03414}{ [arXiv:1507.03414 [hep-ex]]}.



\bibitem{Aaij:2016ymb}
R.~Aaij \textit{et al.} [LHCb],
\href{https://journals.aps.org/prl/abstract/10.1103/PhysRevLett.117.082003}{Phys. Rev. Lett. \textbf{117}, no.8, 082003 (2016)}
\href{https://arxiv.org/abs/1606.06999}{[arXiv:1606.06999 [hep-ex]]}.



\bibitem{Aaij:2019vzc}
R.~Aaij \textit{et al.} [LHCb],
\href{https://doi.org/10.1103/PhysRevLett.122.222001}{Phys. Rev. Lett. \textbf{122}, no.22, 222001 (2019)}
\href{https://arxiv.org/abs/1904.03947}{[arXiv:1904.03947 [hep-ex]]}.



\bibitem{LHCb:2020jpq}
R.~Aaij \textit{et al.} [LHCb],
\href{https://www.sciencedirect.com/science/article/pii/S2095927321001717}{Sci. Bull. \textbf{66}, 1278-1287 (2021)}
\href{https://arxiv.org/abs/2012.10380}{[arXiv:2012.10380 [hep-ex]]}.


\bibitem{LHCb:2021chn}
R.~Aaij \textit{et al.} [LHCb],
\href{https://arxiv.org/abs/2108.04720}{[arXiv:2108.04720 [hep-ex]]}.

\bibitem{Chen:2016heh}
R.~Chen, X.~Liu and S.~L.~Zhu,
\href{https://www.sciencedirect.com/science/article/pii/S0375947416300616?via%3Dihub}{Nucl. Phys. A \textbf{954}, 406-421 (2016)}
\href{https://arxiv.org/abs/1601.03233}{[arXiv:1601.03233 [hep-ph]]}.




\bibitem{Chen:2016otp}
H.~X.~Chen, E.~L.~Cui, W.~Chen, X.~Liu, T.~G.~Steele and S.~L.~Zhu,
\href{https://link.springer.com/article/10.1140%2Fepjc%2Fs10052-016-4438-5}{Eur. Phys. J. C \textbf{76}, no.10, 572 (2016)}
\href{https://arxiv.org/abs/1602.02433}{[arXiv:1602.02433 [hep-ph]]}.



\bibitem{Azizi:2016dhy} 
  K.~Azizi, Y.~Sarac and H.~Sundu,
\href{https://doi.org/10.1103/PhysRevD.95.094016}{  Phys.\ Rev.\ D {\bf 95}, no. 9, 094016 (2017)}
 \href{https://arxiv.org/abs/1612.07479}{ [arXiv:1612.07479 [hep-ph]]}.  
 
 
   \bibitem{Azizi:2018bdv}
K.~Azizi, Y.~Sarac and H.~Sundu,
\href{https://doi.org/10.1016/j.physletb.2018.06.022}{Phys. Lett. B \textbf{782}, 694-701 (2018)}
\href{https://arxiv.org/abs/1802.01384}{[arXiv:1802.01384 [hep-ph]]}.



\bibitem{Azizi:2020ogm}
K.~Azizi, Y.~Sarac and H.~Sundu,
\href{https://iopscience.iop.org/article/10.1088/1674-1137/abe8ce}{Chin. Phys. C \textbf{45}, no.5, 053103 (2021)}
\href{https://arxiv.org/abs/2011.05828}{[arXiv:2011.05828 [hep-ph]]}.




\bibitem{He:2019ify}
J.~He,
\href{https://link.springer.com/article/10.1140%2Fepjc%2Fs10052-019-6906-1}{Eur. Phys. J. C \textbf{79}, no.5, 393 (2019)}
\href{https://arxiv.org/abs/1903.11872}{[arXiv:1903.11872 [hep-ph]]}.


\bibitem{Zhu:2021lhd}
J.~T.~Zhu, L.~Q.~Song and J.~He,
\href{https://journals.aps.org/prd/abstract/10.1103/PhysRevD.103.074007}{Phys. Rev. D \textbf{103}, no.7, 074007 (2021)}
\href{https://arxiv.org/abs/2101.12441}{[arXiv:2101.12441 [hep-ph]]}.

\bibitem{Liu:2019tjn}
M.~Z.~Liu, Y.~W.~Pan, F.~Z.~Peng, M.~S\'anchez S\'anchez, L.~S.~Geng, A.~Hosaka and M.~Pavon Valderrama,
\href{https://journals.aps.org/prl/abstract/10.1103/PhysRevLett.122.242001}{Phys. Rev. Lett. \textbf{122}, no.24, 242001 (2019)}
\href{https://arxiv.org/abs/1903.11560}{[arXiv:1903.11560 [hep-ph]]}.

\bibitem{Xiao:2019mvs}
C.~J.~Xiao, Y.~Huang, Y.~B.~Dong, L.~S.~Geng and D.~Y.~Chen,
\href{https://journals.aps.org/prd/abstract/10.1103/PhysRevD.100.014022}{Phys. Rev. D \textbf{100}, no.1, 014022 (2019)}
\href{https://arxiv.org/abs/1904.00872}{[arXiv:1904.00872 [hep-ph]]}.


\bibitem{Lu:2016nnt}
Q.~F.~L\"u and Y.~B.~Dong,
\href{https://journals.aps.org/prd/abstract/10.1103/PhysRevD.93.074020}{Phys. Rev. D \textbf{93}, no.7, 074020 (2016)}
\href{https://arxiv.org/abs/1603.00559}{[arXiv:1603.00559 [hep-ph]]}.




\bibitem{Hu:2021nvs}
X.~Hu and J.~Ping,
\href{https://arxiv.org/abs/2109.09972v3}{[arXiv:2109.09972 [hep-ph]]}.


\bibitem{Meng:2019dba}
L.~Meng, B.~Wang, G.~J.~Wang and S.~L.~Zhu,
\href{https://arxiv.org/abs/1911.09250}{[arXiv:1911.09250 [hep-ph]]}.


\bibitem{Chen:2015loa}
  R.~Chen, X.~Liu, X.~Q.~Li and S.~L.~Zhu,
\href{https://doi.org/10.1103/PhysRevLett.115.132002}{  Phys.\ Rev.\ Lett.\  {\bf 115}, no. 13, 132002 (2015)}
\href{https://arxiv.org/abs/1507.03704}{  [arXiv:1507.03704 [hep-ph]]}.

\bibitem{Chen:2015moa} 
  H.~X.~Chen, W.~Chen, X.~Liu, T.~G.~Steele and S.~L.~Zhu,
 \href{https://doi.org/10.1103/PhysRevLett.115.172001}{ Phys.\ Rev.\ Lett.\  {\bf 115}, no. 17, 172001 (2015)}
 \href{https://arxiv.org/abs/1507.03717}{ [arXiv:1507.03717 [hep-ph]]}.

  \bibitem{He:2015cea} 
  J.~He,
 \href{https://doi.org/10.1016/j.physletb.2015.12.071}{ Phys.\ Lett.\ B {\bf 753}, 547 (2016)}
\href{https://arxiv.org/abs/1507.05200}{  [arXiv:1507.05200 [hep-ph]]}.
  
\bibitem{Meissner:2015mza}
  U.~G.~Mei\ss ner and J.~A.~Oller,
 \href{https://doi.org/10.1016/j.physletb.2015.10.015}{ Phys.\ Lett.\ B {\bf 751}, 59 (2015)}
\href{https://arxiv.org/abs/1507.07478}{ [arXiv:1507.07478 [hep-ph]]}.  
  




\bibitem{Roca:2015dva} 
  L.~Roca, J.~Nieves and E.~Oset,
\href{https://doi.org/10.1103/PhysRevD.92.094003}{  Phys.\ Rev.\ D {\bf 92}, no. 9, 094003 (2015)}
\href{https://arxiv.org/abs/1507.04249}{  [arXiv:1507.04249 [hep-ph]]}.
  
  



\bibitem{Chen:2020opr}
H.~X.~Chen,
\href{https://arxiv.org/abs/2011.07187}{[arXiv:2011.07187 [hep-ph]]}.


\bibitem{Yan:2021nio}
M.~J.~Yan, F.~Z.~Peng, M.~S.~S\'anchez and M.~P.~Valderrama,
\href{https://arxiv.org/abs/2108.05306}{[arXiv:2108.05306 [hep-ph]]}.


\bibitem{Wu:2021caw}
Q.~Wu, D.~Y.~Chen and R.~Ji,
\href{https://iopscience.iop.org/article/10.1088/0256-307X/38/7/071301}{Chin. Phys. Lett. \textbf{38}, no.7, 071301 (2021)}
\href{https://arxiv.org/abs/2103.05257}{[arXiv:2103.05257 [hep-ph]]}.


\bibitem{Chen:2020uif}
H.~X.~Chen, W.~Chen, X.~Liu and X.~H.~Liu,
\href{https://dx.doi.org/10.1140/epjc/s10052-021-09196-4}{Eur. Phys. J. C \textbf{81}, no.5, 409 (2021)}
\href{https://arxiv.org/abs/2011.01079}{[arXiv:2011.01079 [hep-ph]]}.


\bibitem{Phumphan:2021tta}
K.~Phumphan, W.~Ruangyoo, C.~C.~Chen, A.~Limphirat and Y.~Yan,
\href{https://arxiv.org/abs/2105.03150}{[arXiv:2105.03150 [hep-ph]]}.


\bibitem{Du:2021fmf}
M.~L.~Du, V.~Baru, F.~K.~Guo, C.~Hanhart, U.~G.~Mei\ss{}ner, J.~A.~Oller and Q.~Wang,
\href{https://link.springer.com/content/pdf/10.1007/JHEP08(2021)157.pdf}{JHEP \textbf{08}, 157 (2021)}
\href{https://arxiv.org/abs/2102.07159}{[arXiv:2102.07159 [hep-ph]]}.


\bibitem{Lu:2021irg}
J.~X.~Lu, M.~Z.~Liu, R.~X.~Shi and L.~S.~Geng,
\href{https://journals.aps.org/prd/abstract/10.1103/PhysRevD.104.034022}{Phys. Rev. D \textbf{104}, no.3, 034022 (2021)}
\href{https://arxiv.org/abs/2104.10303}{[arXiv:2104.10303 [hep-ph]]}.


\bibitem{Gao:2021hmv}
F.~Gao and H.~S.~Li,
\href{https://arxiv.org/abs/2112.01823}{[arXiv:2112.01823 [hep-ph]]}.

\bibitem{Yalikun:2021dpk}
N.~Yalikun and B.~S.~Zou,
\href{https://arxiv.org/abs/2112.06426}{[arXiv:2112.06426 [hep-ph]]}.


\bibitem{Chen:2021obo}
C.~h.~Chen, Y.~L.~Xie, H.~g.~Xu, Z.~Zhang, D.~M.~Zhou, Z.~L.~She and G.~Chen,
\href{https://arxiv.org/abs/2111.03241}{[arXiv:2111.03241 [hep-ph]]}.


\bibitem{Wang:2015epa}
Z.~G.~Wang,
\href{https://link.springer.com/article/10.1140%2Fepjc%2Fs10052-016-3920-4}{Eur. Phys. J. C \textbf{76}, no.2, 70 (2016)}
\href{https://arxiv.org/abs/1508.01468}{[arXiv:1508.01468 [hep-ph]]}.



\bibitem{Maiani:2015vwa} 
  L.~Maiani, A.~D.~Polosa and V.~Riquer,
 \href{https://doi.org/10.1016/j.physletb.2015.08.008}{ Phys.\ Lett.\ B {\bf 749}, 289 (2015)}
 \href{https://arxiv.org/abs/1507.04980}{ [arXiv:1507.04980 [hep-ph]]}.
 
 
 \bibitem{Giannuzzi:2019esi}
F.~Giannuzzi,
\href{https://journals.aps.org/prd/abstract/10.1103/PhysRevD.99.094006}{Phys. Rev. D \textbf{99}, no.9, 094006 (2019)}
\href{https://arxiv.org/abs/1903.04430}{[arXiv:1903.04430 [hep-ph]]}.


 \bibitem{Li:2015gta} 
  G.~N.~Li, X.~G.~He and M.~He,
 \href{https://doi.org/10.1007/JHEP12(2015)128}{ JHEP {\bf 12}, 128 (2015)}
  \href{https://arxiv.org/abs/1507.08252}{[arXiv:1507.08252 [hep-ph]]}.
  
  

  \bibitem{Lebed:2015tna} 
  R.~F.~Lebed,
 \href{https://doi.org/10.1016/j.physletb.2015.08.032}{ Phys.\ Lett.\ B {\bf 749}, 454 (2015)}
 \href{https://arxiv.org/abs/1507.05867}{ [arXiv:1507.05867 [hep-ph]]}.
  
  
 
 
  
  
 \bibitem{Anisovich:2015cia}
  V.~V.~Anisovich, M.~A.~Matveev, J.~Nyiri, A.~V.~Sarantsev and A.~N.~Semenova,
 \href{https://arxiv.org/abs/1507.07652}{ arXiv:1507.07652 [hep-ph]}.


  
  

\bibitem{Wang:2015ava}
  Z.~G.~Wang and T.~Huang,
\href{https://doi.org/10.1140/epjc/s10052-016-3880-8}{  Eur.\ Phys.\ J.\ C {\bf 76}, no. 1, 43 (2016)}
\href{https://arxiv.org/abs/1508.04189}{  [arXiv:1508.04189 [hep-ph]]}.



  
  
  \bibitem{Wang:2015ixb}
Z.~G.~Wang,
\href{https://doi.org/10.1016/j.nuclphysb.2016.09.009}{Nucl. Phys. B \textbf{913}, 163-208 (2016)}
\href{https://arxiv.org/abs/1512.04763}{[arXiv:1512.04763 [hep-ph]]}.
  
 
  
 \bibitem{Ghosh:2015ksa}
R.~Ghosh, A.~Bhattacharya and B.~Chakrabarti,
\href{https://doi.org/10.1134/S1547477117040100}{Phys. Part. Nucl. Lett. \textbf{14}, no.4, 550-552 (2017)}
\href{https://arxiv.org/abs/1508.00356}{[arXiv:1508.00356 [hep-ph]]}.  
  
  

\bibitem{Wang:2015wsa}
Z.~G.~Wang,
\href{https://link.springer.com/article/10.1140/epjc/s10052-016-3983-2}{Eur. Phys. J. C \textbf{76}, no.3, 142 (2016)}
\href{https://arxiv.org/abs/1509.06436}{[arXiv:1509.06436 [hep-ph]]}.  
  
 
  

\bibitem{Zhang:2017mmw}
J.~X.~Zhang, Z.~G.~Wang and Z.~Y.~Di,
\href{https://www.actaphys.uj.edu.pl/index_n.php?I=R&V=48&N=11#2013}{Acta Phys. Polon. B \textbf{48}, 2013 (2017)}
\href{https://arxiv.org/abs/1711.10728}{[arXiv:1711.10728 [hep-ph]]}.  
  
  
\bibitem{Wang:2019got}
Z.~G.~Wang,
\href{https://www.worldscientific.com/doi/abs/10.1142/S0217751X20500037}{Int. J. Mod. Phys. A \textbf{35}, no.01, 2050003 (2020)}
\href{https://arxiv.org/abs/1905.02892}{[arXiv:1905.02892 [hep-ph]]}.  
  
  
  
\bibitem{Wang:2020rdh}
Z.~G.~Wang, H.~J.~Wang and Q.~Xin,
\href{http://cpc.ihep.ac.cn/article/doi/10.1088/1674-1137/abf13b}{Chin. Phys. C \textbf{45}, 063104 (2021)}
\href{https://arxiv.org/abs/2005.00535v3}{[arXiv:2005.00535 [hep-ph]]}.  


\bibitem{Ali:2020vee}
A.~Ali, I.~Ahmed, M.~J.~Aslam, A.~Parkhomenko and A.~Rehman,
\href{https://pos.sissa.it/390/527/}{PoS \textbf{ICHEP2020}, 527 (2021)}
\href{https://arxiv.org/abs/2012.07760v1}{[arXiv:2012.07760 [hep-ph]]}.


  
  \bibitem{Wang:2016dzu}
  G.~J.~Wang, R.~Chen, L.~Ma, X.~Liu and S.~L.~Zhu,
 \href{https://doi.org/10.1103/PhysRevD.94.094018}{ Phys.\ Rev.\ D {\bf 94}, no. 9, 094018 (2016)}
 \href{https://arxiv.org/abs/1605.01337}{ [arXiv:1605.01337 [hep-ph]]}. 
 
  
\bibitem{Azizi:2021utt}
K.~Azizi, Y.~Sarac and H.~Sundu,
\href{https://journals.aps.org/prd/abstract/10.1103/PhysRevD.103.094033}{Phys. Rev. D \textbf{103}, no.9, 094033 (2021)}
\href{https://arxiv.org/abs/2101.07850}{[arXiv:2101.07850 [hep-ph]]}.


\bibitem{Anisovich:2015zqa}
V.~V.~Anisovich, M.~A.~Matveev, J.~Nyiri, A.~V.~Sarantsev and A.~N.~Semenova,
\href{https://www.worldscientific.com/doi/abs/10.1142/S0217751X15501900}{Int. J. Mod. Phys. A \textbf{30}, no.32, 1550190 (2015)}
\href{https://arxiv.org/abs/1509.04898}{[arXiv:1509.04898 [hep-ph]]}.



\bibitem{Wang:2020eep}
Z.~G.~Wang,
\href{https://www.worldscientific.com/doi/abs/10.1142/S0217751X21500718}{Int. J. Mod. Phys. A \textbf{36}, no.10, 2150071 (2021)}
\href{https://arxiv.org/abs/2011.05102}{[arXiv:2011.05102 [hep-ph]]}.




 \bibitem{Zhu:2015bba}
  R.~Zhu and C.~F.~Qiao,
 \href{https://doi.org/10.1016/j.physletb.2016.03.022}{ Phys.\ Lett.\ B {\bf 756}, 259 (2016)}
 \href{https://arxiv.org/abs/1510.08693}{ [arXiv:1510.08693 [hep-ph]]}. 
 
 
 \bibitem{Guo1}
F.~K.~Guo, U.~G.~Mei\ss ner, W.~Wang and Z.~Yang,
\href{https://doi.org/10.1103/PhysRevD.92.071502}{Phys. Rev. D \textbf{92}, no.7, 071502 (2015)}
\href{https://arxiv.org/abs/1507.04950}{[arXiv:1507.04950 [hep-ph]]}.



\bibitem{Guo2}
F.~K.~Guo, U.~G.~Mei\ss ner, J.~Nieves and Z.~Yang,
\href{https://doi.org/10.1140/epja/i2016-16318-4}{Eur. Phys. J. A \textbf{52}, no.10, 318 (2016)}
\href{https://arxiv.org/abs/1605.05113}{[arXiv:1605.05113 [hep-ph]]}.

\bibitem{Mikhasenko:2015vca} 
  M.~Mikhasenko,
 \href{https://arxiv.org/abs/1507.06552}{ arXiv:1507.06552 [hep-ph]}.


\bibitem{Liu1}
X.~H.~Liu, Q.~Wang and Q.~Zhao,
\href{https://doi.org/10.1016/j.physletb.2016.03.089}{Phys. Lett. B \textbf{757}, 231-236 (2016)}
\href{https://arxiv.org/abs/1507.05359}{[arXiv:1507.05359 [hep-ph]]}.  


\bibitem{Bayar:2016ftu}
M.~Bayar, F.~Aceti, F.~K.~Guo and E.~Oset,
\href{https://journals.aps.org/prd/abstract/10.1103/PhysRevD.94.074039}{Phys. Rev. D \textbf{94}, no.7, 074039 (2016)}
\href{https://arxiv.org/abs/1609.04133}{[arXiv:1609.04133 [hep-ph]]}.


\bibitem{Nakamura:2021qvy}
S.~X.~Nakamura,
\href{https://doi.org/10.1103/PhysRevD.103.L111503}{Phys. Rev. D \textbf{103} (2021), 111503}
\href{https://arxiv.org/abs/2103.06817}{[arXiv:2103.06817 [hep-ph]]}.




\bibitem{Nakamura:2021dix}
S.~X.~Nakamura, A.~Hosaka and Y.~Yamaguchi,
\href{https://journals.aps.org/prd/abstract/10.1103/PhysRevD.104.L091503}{Phys. Rev. D \textbf{104}, no.9, L091503 (2021)}
\href{https://arxiv.org/abs/2109.15235}{[arXiv:2109.15235 [hep-ph]]}.


\bibitem{Liu:2020cmw}
X.~Liu, H.~Huang and J.~Ping,
\href{https://journals.aps.org/prc/abstract/10.1103/PhysRevC.102.025204}{Phys. Rev. C \textbf{102} (2020) no.2, 025204}
\href{https://arxiv.org/abs/2005.09646}{[arXiv:2005.09646 [hep-ph]]}.
  



\bibitem{Chen:2015sxa}
H.~X.~Chen, L.~S.~Geng, W.~H.~Liang, E.~Oset, E.~Wang and J.~J.~Xie,
\href{https://doi.org/10.1103/PhysRevC.93.065203}{Phys. Rev. C \textbf{93}, no.6, 065203 (2016)}
\href{https://arxiv.org/abs/1510.01803}{[arXiv:1510.01803 [hep-ph]]}.

\bibitem{Feijoo:2015kts}
A.~Feijoo, V.~K.~Magas, A.~Ramos and E.~Oset,
\href{https://doi.org/10.1140/epjc/s10052-016-4302-7}{Eur. Phys. J. C \textbf{76}, no.8, 446 (2016)}
\href{https://arxiv.org/abs/1512.08152}{[arXiv:1512.08152 [hep-ph]]}.





\bibitem{Irie:2017qai}
Y.~Irie, M.~Oka and S.~Yasui,
\href{https://doi.org/10.1103/PhysRevD.97.034006}{Phys. Rev. D \textbf{97}, no.3, 034006 (2018)}
\href{https://arxiv.org/abs/1707.04544}{[arXiv:1707.04544 [hep-ph]]}.


\bibitem{Chen:2016ryt}
R.~Chen, J.~He and X.~Liu,
\href{https://dx.doi.org/10.1088/1674-1137/41/10/103105}{Chin. Phys. C \textbf{41}, no.10, 103105 (2017)}
\href{https://arxiv.org/abs/1609.03235}{[arXiv:1609.03235 [hep-ph]]}.


\bibitem{Zhang:2020cdi}
Q.~Zhang, B.~R.~He and J.~L.~Ping,
\href{https://arxiv.org/abs/2006.01042}{[arXiv:2006.01042 [hep-ph]]}.

\bibitem{Paryev:2020jkp}
E.~Y.~Paryev,
\href{https://arxiv.org/abs/2007.01172}{[arXiv:2007.01172 [nucl-th]]}.


\bibitem{Gutsche:2019mkg}
T.~Gutsche and V.~E.~Lyubovitskij,
\href{https://doi.org/10.1103/PhysRevD.100.094031}{Phys. Rev. D \textbf{100}, no.9, 094031 (2019)}
\href{https://arxiv.org/abs/1910.03984}{[arXiv:1910.03984 [hep-ph]]}.


\bibitem{Azizi:2017bgs}
K.~Azizi, Y.~Sarac and H.~Sundu,
\href{https://doi.org/10.1103/PhysRevD.96.094030}{Phys. Rev. D \textbf{96}, no.9, 094030 (2017)}
\href{https://arxiv.org/abs/1707.01248}{[arXiv:1707.01248 [hep-ph]]}.

\bibitem{Azizi:2018dva}
K.~Azizi, Y.~Sarac and H.~Sundu,
\href{https://doi.org/10.1103/PhysRevD.98.054002}{Phys. Rev. D \textbf{98}, no.5, 054002 (2018)}
\href{https://arxiv.org/abs/1805.06734}{[arXiv:1805.06734 [hep-ph]]}.



\bibitem{Cao:2019gqo}
X.~Cao, F.~K.~Guo, Y.~T.~Liang, J.~J.~Wu, J.~J.~Xie, Y.~P.~Xie, Z.~Yang and B.~S.~Zou,
\href{https://doi.org/10.1103/PhysRevD.101.074010}{Phys. Rev. D \textbf{101}, no.7, 074010 (2020)}
\href{https://arxiv.org/abs/1912.12054}{[arXiv:1912.12054 [hep-ph]]}.





\bibitem{Zhang:2020vpz}
J.~R.~Zhang,
\href{https://journals.aps.org/prd/abstract/10.1103/PhysRevD.103.074016}{Phys. Rev. D \textbf{103}, no.7, 074016 (2021)}
\href{https://arxiv.org/abs/2011.04594v2}{[arXiv:2011.04594 [hep-ph]]}.




\bibitem{Xie:2020ckr}
Y.~P.~Xie and V.~P.~Goncalves,
\href{https://www.sciencedirect.com/science/article/pii/S0370269321000617?via%3Dihub}{Phys. Lett. B \textbf{814}, 136121 (2021)}
\href{https://arxiv.org/abs/2012.07501}{[arXiv:2012.07501 [hep-ph]]}.



\bibitem{Wang:2020bjt}
F.~L.~Wang, R.~Chen and X.~Liu,
\href{https://journals.aps.org/prd/abstract/10.1103/PhysRevD.103.034014}{Phys. Rev. D \textbf{103}, no.3, 034014 (2021)}
\href{https://arxiv.org/abs/2011.14296}{[arXiv:2011.14296 [hep-ph]]}.



\bibitem{Meng:2019fan}
Q.~Meng, E.~Hiyama, K.~U.~Can, P.~Gubler, M.~Oka, A.~Hosaka and H.~Zong,
\href{https://www.sciencedirect.com/science/article/pii/S0370269319307506?via%3Dihub}{Phys. Lett. B \textbf{798}, 135028 (2019)}
\href{https://arxiv.org/abs/1907.00144}{[arXiv:1907.00144 [nucl-th]]}.


\bibitem{Liu:2021tpq}
Y.~Liu, M.~A.~Nowak and I.~Zahed,
\href{https://arxiv.org/abs/2108.04334}{[arXiv:2108.04334 [hep-ph]]}.


\bibitem{Wang:2019nvm}
B.~Wang, L.~Meng and S.~L.~Zhu,
\href{https://journals.aps.org/prd/abstract/10.1103/PhysRevD.101.034018}{Phys. Rev. D \textbf{101}, no.3, 034018 (2020)}
\href{https://arxiv.org/abs/1912.12592}{[arXiv:1912.12592 [hep-ph]]}.





\bibitem{Liu:2021ixf}
Y.~Liu, M.~A.~Nowak and I.~Zahed,
\href{https://journals.aps.org/prd/abstract/10.1103/PhysRevD.104.114022}{Phys. Rev. D \textbf{104}, no.11, 114022 (2021)}
\href{https://arxiv.org/abs/2108.07074}{[arXiv:2108.07074 [hep-ph]]}.




\bibitem{Giron:2021fnl}
J.~F.~Giron and R.~F.~Lebed,
\href{https://journals.aps.org/prd/abstract/10.1103/PhysRevD.104.114028}{Phys. Rev. D \textbf{104}, no.11, 114028 (2021)}
\href{https://arxiv.org/abs/2110.05557}{[arXiv:2110.05557 [hep-ph]]}.

\bibitem{Santopinto:2016pkp}
E.~Santopinto and A.~Giachino,
\href{https://journals.aps.org/prd/abstract/10.1103/PhysRevD.96.014014}{Phys. Rev. D \textbf{96}, no.1, 014014 (2017)}
\href{https://arxiv.org/abs/1604.03769}{[arXiv:1604.03769 [hep-ph]]}.


\bibitem{Xiao:2019gjd}
C.~W.~Xiao, J.~Nieves and E.~Oset,
\href{https://www.sciencedirect.com/science/article/pii/S0370269319307737?via%3Dihub}{Phys. Lett. B \textbf{799}, 135051 (2019)}
\href{https://arxiv.org/abs/1906.09010}{[arXiv:1906.09010 [hep-ph]]}.





\bibitem{Wang:2015jsa}
Q.~Wang, X.~H.~Liu and Q.~Zhao,
\href{https://journals.aps.org/prd/abstract/10.1103/PhysRevD.92.034022}{Phys. Rev. D \textbf{92}, 034022 (2015)}
\href{https://arxiv.org/abs/1508.00339}{[arXiv:1508.00339 [hep-ph]]}.


\bibitem{Kubarovsky:2015aaa}
V.~Kubarovsky and M.~B.~Voloshin,
\href{https://journals.aps.org/prd/abstract/10.1103/PhysRevD.92.031502}{Phys. Rev. D \textbf{92}, no.3, 031502 (2015)}
\href{https://arxiv.org/abs/1508.00888}{[arXiv:1508.00888 [hep-ph]]}.

\bibitem{Karliner:2015voa}
M.~Karliner and J.~L.~Rosner,
\href{https://www.sciencedirect.com/science/article/pii/S0370269315009247?via%3Dihub}{Phys. Lett. B \textbf{752}, 329-332 (2016)}
\href{https://arxiv.org/abs/1508.01496}{[arXiv:1508.01496 [hep-ph]]}.


\bibitem{Cheng:2015cca}
H.~Y.~Cheng and C.~K.~Chua,
\href{https://journals.aps.org/prd/abstract/10.1103/PhysRevD.92.096009}{Phys. Rev. D \textbf{92}, no.9, 096009 (2015)}
\href{https://arxiv.org/abs/1509.03708}{[arXiv:1509.03708 [hep-ph]]}.


\bibitem{Liu:2020ajv}
W.~Y.~Liu, W.~Hao, G.~Y.~Wang, Y.~Y.~Wang, E.~Wang and D.~M.~Li,
\href{https://journals.aps.org/prd/abstract/10.1103/PhysRevD.103.034019}{Phys. Rev. D \textbf{103}, no.3, 034019 (2021)}
\href{https://arxiv.org/abs/2012.01804v2}{[arXiv:2012.01804 [hep-ph]]}.




\bibitem{Yang:2021pio}
F.~Yang, Y.~Huang and H.~Q.~Zhu,
\href{https://dx.doi.org/10.1007/s11433-021-1796-0}{Sci. China Phys. Mech. Astron. \textbf{64}, no.12, 121011 (2021)}
\href{https://arxiv.org/abs/2107.13267}{[arXiv:2107.13267 [hep-ph]]}.


\bibitem{Stancu:2021rro}
F.~Stancu,
\href{https://journals.aps.org/prd/abstract/10.1103/PhysRevD.104.054050}{Phys. Rev. D \textbf{104}, no.5, 054050 (2021)}
\href{https://arxiv.org/abs/2108.05841}{[arXiv:2108.05841 [hep-ph]]}.

\bibitem{Dong:2020nwk}
Y.~Dong, P.~Shen, F.~Huang and Z.~Zhang,
\href{https://link.springer.com/article/10.1140%2Fepjc%2Fs10052-020-7890-1}{Eur. Phys. J. C \textbf{80}, no.4, 341 (2020)}
\href{https://arxiv.org/abs/2002.08051}{[arXiv:2002.08051 [hep-ph]]}.


\bibitem{Voloshin:2019aut}
M.~B.~Voloshin,
\href{https://journals.aps.org/prd/abstract/10.1103/PhysRevD.100.034020}{Phys. Rev. D \textbf{100}, no.3, 034020 (2019)}
\href{https://arxiv.org/abs/1907.01476}{[arXiv:1907.01476 [hep-ph]]}.



\bibitem{Ling:2021lmq}
X.~Z.~Ling, J.~X.~Lu, M.~Z.~Liu and L.~S.~Geng,
\href{https://journals.aps.org/prd/abstract/10.1103/PhysRevD.104.074022}{Phys. Rev. D \textbf{104}, no.7, 074022 (2021)}
\href{https://arxiv.org/abs/2106.12250}{[arXiv:2106.12250 [hep-ph]]}.



\bibitem{Xing:2021yid}
Y.~Xing and Y.~Niu,
\href{https://epjc.epj.org/articles/epjc/abs/2021/11/10052_2021_Article_9730/10052_2021_Article_9730.html}{Eur. Phys. J. C \textbf{81}, no.11, 978 (2021)}
\href{https://arxiv.org/abs/2106.09939}{[arXiv:2106.09939 [hep-ph]]}.


\bibitem{Cheng:2021gca}
C.~Cheng, F.~Yang and Y.~Huang,
\href{https://arxiv.org/abs/2110.04746}{[arXiv:2110.04746 [hep-ph]]}.


\bibitem{Wang:2019hyc}
Z.~G.~Wang and X.~Wang,
\href{https://iopscience.iop.org/article/10.1088/1674-1137/ababf7}{Chin. Phys. C \textbf{44}, 103102 (2020)}
\href{https://arxiv.org/abs/1907.04582v3}{[arXiv:1907.04582 [hep-ph]]}.


\bibitem{Xu:2019zme}
Y.~J.~Xu, C.~Y.~Cui, Y.~L.~Liu and M.~Q.~Huang,
\href{https://journals.aps.org/prd/abstract/10.1103/PhysRevD.102.034028}{Phys. Rev. D \textbf{102}, no.3, 034028 (2020)}
\href{https://arxiv.org/abs/1907.05097}{[arXiv:1907.05097 [hep-ph]]}.


\bibitem{Liu:2021ojf}
J.~Liu, D.~Y.~Chen and J.~He,
\href{https://link.springer.com/article/10.1140%2Fepjc%2Fs10052-021-09766-6}{Eur. Phys. J. C \textbf{81}, no.11, 965 (2021)}
\href{https://arxiv.org/abs/2108.00148}{[arXiv:2108.00148 [hep-ph]]}.



\bibitem{Huang:2021ave}
Y.~Huang and H.~Q.~Zhu,
\href{https://journals.aps.org/prd/abstract/10.1103/PhysRevD.104.056027}{Phys. Rev. D \textbf{104}, no.5, 056027 (2021)}
\href{https://arxiv.org/abs/2107.03773}{[arXiv:2107.03773 [hep-ph]]}.



\bibitem{Zhu:2020vto}
J.~T.~Zhu, S.~Y.~Kong, Y.~Liu and J.~He,
\href{https://link.springer.com/article/10.1140%2Fepjc%2Fs10052-020-8410-z}{Eur. Phys. J. C \textbf{80}, no.11, 1016 (2020)}
\href{https://arxiv.org/abs/2007.07596}{[arXiv:2007.07596 [hep-ph]]}.



\bibitem{Ferretti:2018ojb}
J.~Ferretti, E.~Santopinto, M.~Naeem Anwar and M.~A.~Bedolla,
\href{https://www.sciencedirect.com/science/article/pii/S0370269318307482?via%3Dihub}{Phys. Lett. B \textbf{789}, 562-567 (2019)}
\href{https://arxiv.org/abs/1807.01207}{[arXiv:1807.01207 [hep-ph]]}.

\bibitem{Shimizu:2016rrd}
Y.~Shimizu, D.~Suenaga and M.~Harada,
\href{https://journals.aps.org/prd/abstract/10.1103/PhysRevD.93.114003}{Phys. Rev. D \textbf{93}, no.11, 114003 (2016)}
\href{https://arxiv.org/abs/1603.02376}{[arXiv:1603.02376 [hep-ph]]}.




\bibitem{Liu:2017xzo}
Y.~Liu and I.~Zahed,
\href{https://journals.aps.org/prd/abstract/10.1103/PhysRevD.95.116012}{Phys. Rev. D \textbf{95}, no.11, 116012 (2017)}
\href{https://arxiv.org/abs/1704.03412}{[arXiv:1704.03412 [hep-ph]]}.



\bibitem{Wang:2021xao}
Z.~G.~Wang,
\href{https://www.sciencedirect.com/science/article/pii/S0550321321002765?via%3Dihub}{Nucl. Phys. B \textbf{973}, 115579 (2021)}
\href{https://arxiv.org/abs/2104.12090}{[arXiv:2104.12090 [hep-ph]]}.


\bibitem{An:2020jix}
H.~T.~An, K.~Chen, Z.~W.~Liu and X.~Liu,
\href{https://journals.aps.org/prd/abstract/10.1103/PhysRevD.103.074006}{Phys. Rev. D \textbf{103}, no.7, 074006 (2021)}
\href{https://arxiv.org/abs/2012.12459}{[arXiv:2012.12459 [hep-ph]]}.


\bibitem{Yan:2021glh}
Y.~Yan, Y.~Wu, X.~Hu, H.~Huang and J.~Ping,
\href{https://arxiv.org/abs/2110.10853}{[arXiv:2110.10853 [hep-ph]]}.




\bibitem{Ferretti:2020ewe}
J.~Ferretti and E.~Santopinto,
\href{https://link.springer.com/article/10.1007%2FJHEP04%282020%29119}{JHEP \textbf{04}, 119 (2020)}
\href{https://arxiv.org/abs/2001.01067v3}{[arXiv:2001.01067 [hep-ph]]}.

\bibitem{Ferretti:2021zis}
J.~Ferretti and E.~Santopinto,
\href{https://arxiv.org/abs/2111.08650}{[arXiv:2111.08650 [hep-ph]]}.


\bibitem{Wang:2021hql}
F.~L.~Wang, X.~D.~Yang, R.~Chen and X.~Liu,
\href{https://journals.aps.org/prd/abstract/10.1103/PhysRevD.103.054025}{Phys. Rev. D \textbf{103}, no.5, 054025 (2021)}
\href{https://arxiv.org/abs/2101.11200}{[arXiv:2101.11200 [hep-ph]]}.




\bibitem{Shifman:1978bx}
  M.~A.~Shifman, A.~I.~Vainshtein and V.~I.~Zakharov,
 \href{https://doi.org/10.1016/0550-3213(79)90022-1}{ Nucl.\ Phys.\ B {\bf 147}, 385 (1979)}.
  

  
\bibitem{Shifman:1978by}
  M.~A.~Shifman, A.~I.~Vainshtein and V.~I.~Zakharov,
 \href{https://doi.org/10.1016/0550-3213(79)90023-3}{ Nucl.\ Phys.\ B {\bf 147}, 448 (1979)}.
 
 

 \bibitem{Ioffe81} 
  B.~L.~Ioffe,
 \href{https://doi.org/10.1016/0550-3213(81)90259-5}{ Nucl.\ Phys.\ B {\bf 188}, 317 (1981)}
 \href{https://doi.org/10.1016/0550-3213(81)90315-1}{ Erratum: [Nucl.\ Phys.\ B {\bf 191}, 591 (1981)]}.


\bibitem{Zyla:2020zbs}
P.~A.~Zyla \textit{et al.} [Particle Data Group],
\href{https://academic.oup.com/ptep/article/2020/8/083C01/5891211}{PTEP \textbf{2020}, no.8, 083C01 (2020)}.
  





 
 \bibitem{Belyaev:1982sa}
  V.~M.~Belyaev and B.~L.~Ioffe,
  \href{http://www.jetp.ac.ru/cgi-bin/dn/e_056_03_0493.pdf}{Sov. Phys. JETP \textbf{56}, 493-501 (1982)}
  [Zh.\ Eksp.\ Teor.\ Fiz.\  {\bf 83}, 876 (1982)].
  

  
   \bibitem{Belyaev:1982cd}
  V.~M.~Belyaev and B.~L.~Ioffe,
  \href{http://www.jetp.ac.ru/cgi-bin/dn/e_057_04_0716.pdf}{Sov.\ Phys.\ JETP {\bf 57}, 716 (1983)}
  [Zh.\ Eksp.\ Teor.\ Fiz.\  {\bf 84}, 1236 (1983)].




\bibitem{Wang:2007yt}
Z.~G.~Wang,
\href{https://journals.aps.org/prd/abstract/10.1103/PhysRevD.75.054020}{Phys. Rev. D \textbf{75}, 054020 (2007)}
\href{https://arxiv.org/abs/hep-ph/0701176}{[arXiv:hep-ph/0701176 [hep-ph]]}.



\bibitem{Wang:2010fq}
Z.~G.~Wang,
\href{https://link.springer.com/article/10.1140/epjc/s10052-010-1365-8}{Eur. Phys. J. C \textbf{68}, 479-486 (2010)}
\href{https://arxiv.org/abs/1001.1652}{[arXiv:1001.1652 [hep-ph]]}.






\bibitem{Veliev:2011kq}
  E.~V.~Veliev, K.~Azizi, H.~Sundu, G.~Kaya and A.~Turkan,
\href{https://link.springer.com/article/10.1140/epja/i2011-11110-8}{  Eur.\ Phys.\ J.\ A {\bf 47}, 110 (2011)}
 \href{https://arxiv.org/abs/1103.4330}{[arXiv:1103.4330 [hep-ph]]}.
 


\bibitem{Wang:2015mxa}
Z.~G.~Wang,
\href{https://link.springer.com/article/10.1140/epjc/s10052-015-3653-9}{Eur. Phys. J. C \textbf{75}, 427 (2015)}
\href{https://arxiv.org/abs/1506.01993}{[arXiv:1506.01993 [hep-ph]]}.



\end{thebibliography}
\end{document}